\documentclass{aastex}
\usepackage{emulateapj5}

\newcommand{\etal}{et~al.\ }
\newcommand{\CaIIdblt}{{\rm Ca}\kern 0.1em{\sc ii}~$\lambda\lambda 3934, 3969$}
\newcommand{\CIVdblt}{{\rm C}\kern 0.1em{\sc iv}~$\lambda\lambda 1548, 1550$}
\newcommand{\MgIIdblt}{{\rm Mg}\kern 0.1em{\sc ii}~$\lambda\lambda 2796, 2803$}
\newcommand{\NVdblt}{{\rm N}\kern 0.1em{\sc v}~$\lambda\lambda 1238, 1242$}
\newcommand{\SVIdblt}{{\rm S}\kern 0.1em{\sc vi}~$\lambda\lambda 933, 944$}
\newcommand{\OVIdblt}{{\rm O}\kern 0.1em{\sc vi}~$\lambda\lambda 1031, 1037$}
\newcommand{\SiIIdblt}{{\rm Si}\kern 0.1em{\sc ii}~$\lambda\lambda 1190, 1193$}
\newcommand{\SiIVdblt}{{\rm Si}\kern 0.1em{\sc iv}~$\lambda\lambda 1393, 1402$}
\newcommand{\PV}{\hbox{{\rm P}\kern 0.1em{\sc v}}}
\newcommand{\AlI}{\hbox{{\rm Al}\kern 0.1em{\sc i}}}
\newcommand{\AlII}{\hbox{{\rm Al}\kern 0.1em{\sc ii}}}
\newcommand{\AlIII}{{\hbox{\rm Al}\kern 0.1em{\sc iii}}}
\newcommand{\CaII}{\hbox{{\rm Ca}\kern 0.1em{\sc ii}}}
\newcommand{\CII}{\hbox{{\rm C}\kern 0.1em{\sc ii}}}
\newcommand{\CIIe}{\hbox{{\rm C$^{\ast}$}\kern 0.1em{\sc ii}}}
\newcommand{\CIII}{\hbox{{\rm C}\kern 0.1em{\sc iii}}}
\newcommand{\CIV}{\hbox{{\rm C}\kern 0.1em{\sc iv}}}
\newcommand{\CV}{\hbox{{\rm C}\kern 0.1em{\sc v}}}
\newcommand{\HI}{\hbox{{\rm H}\kern 0.1em{\sc i}}}
\newcommand{\HII}{\hbox{{\rm H}\kern 0.1em{\sc ii}}}
\newcommand{\Lya}{\hbox{{\rm Ly}\kern 0.1em$\alpha$}}
\newcommand{\Lyb}{\hbox{{\rm Ly}\kern 0.1em$\beta$}}
\newcommand{\Lyg}{\hbox{{\rm Ly}\kern 0.1em$\gamma$}}
\newcommand{\Lyd}{\hbox{{\rm Ly}\kern 0.1em$\delta$}}
\newcommand{\Lye}{\hbox{{\rm Ly}\kern 0.1em$\epsilon$}}
\newcommand{\Lyphi}{\hbox{{\rm Ly}\kern 0.1em$\phi$}}
\newcommand{\Lyfive}{\hbox{{\rm Ly}\kern 0.1em$5$}}
\newcommand{\Lysix}{\hbox{{\rm Ly}\kern 0.1em$6$}}
\newcommand{\Lyseven}{\hbox{{\rm Ly}\kern 0.1em$7$}}
\newcommand{\Lyeight}{\hbox{{\rm Ly}\kern 0.1em$8$}}
\newcommand{\Lynine}{\hbox{{\rm Ly}\kern 0.1em$9$}}
\newcommand{\Lyten}{\hbox{{\rm Ly}\kern 0.1em$10$}}
\newcommand{\Lyeleven}{\hbox{{\rm Ly}\kern 0.1em$11$}}
\newcommand{\HeI}{\hbox{{\rm He}\kern 0.1em{\sc i}}}
\newcommand{\HeII}{\hbox{{\rm He}\kern 0.1em{\sc ii}}}
\newcommand{\FeI}{\hbox{{\rm Fe}\kern 0.1em{\sc i}}}
\newcommand{\FeII}{\hbox{{\rm Fe}\kern 0.1em{\sc ii}}}
\newcommand{\FeIII}{\hbox{{\rm Fe}\kern 0.1em{\sc iii}}}
\newcommand{\MnII}{\hbox{{\rm Mn}\kern 0.1em{\sc ii}}}
\newcommand{\MgI}{\hbox{{\rm Mg}\kern 0.1em{\sc i}}}
\newcommand{\MgII}{\hbox{{\rm Mg}\kern 0.1em{\sc ii}}}
\newcommand{\MgIII}{\hbox{{\rm Mg}\kern 0.1em{\sc iii}}}
\newcommand{\NI}{\hbox{{\rm N}\kern 0.1em{\sc i}}}
\newcommand{\NII}{\hbox{{\rm N}\kern 0.1em{\sc ii}}}
\newcommand{\NIII}{\hbox{{\rm N}\kern 0.1em{\sc iii}}}
\newcommand{\NV}{\hbox{{\rm N}\kern 0.1em{\sc v}}}
\newcommand{\OVI}{\hbox{{\rm O}\kern 0.1em{\sc vi}}}
\newcommand{\OI}{\hbox{{\rm O}\kern 0.1em{\sc i}}}
\newcommand{\OII}{\hbox{[{\rm O}\kern 0.1em{\sc ii}]}}
\newcommand{\OIV}{\hbox{{\rm O}\kern 0.1em{\sc iv}]}}
\newcommand{\SI}{{\rm S}\kern 0.1em{\sc i}}
\newcommand{\SIV}{{\rm S}\kern 0.1em{\sc iv}}
\newcommand{\SVI}{{\rm S}\kern 0.1em{\sc vi}}
\newcommand{\SiI}{\hbox{{\rm Si}\kern 0.1em{\sc i}}}
\newcommand{\SiII}{\hbox{{\rm Si}\kern 0.1em{\sc ii}}}
\newcommand{\SiIII}{\hbox{{\rm Si}\kern 0.1em{\sc iii}}}
\newcommand{\SiIV}{\hbox{{\rm Si}\kern 0.1em{\sc iv}}}
\newcommand{\SII}{\hbox{{\rm S}\kern 0.1em{\sc ii}}}
\newcommand{\SIII}{\hbox{{\rm S}\kern 0.1em{\sc iii}}}
\newcommand{\NaI}{\hbox{{\rm Na}\kern 0.1em{\sc i}}}
\newcommand{\TiII}{\hbox{{\rm Ti}\kern 0.1em{\sc ii}}}
\newcommand{\kms}{\hbox{km~s$^{-1}$}}
\newcommand{\cmsq}{\hbox{cm$^{-2}$}}
\newcommand{\cc}{\hbox{cm$^{-3}$}}

\shorttitle{VP ANALYSIS OF {\MgII} ABSORBERS}
\shortauthors{CHURCHILL, VOGT, \& CHARLTON.}
\slugcomment{The Astronomical Journal, in press (January 2003)}

\begin{document}

\title{The Physical Conditions of Intermediate Redshift
M\lowercase{g}~{\sc ii} Absorbing Clouds from Voigt Profile
Analysis\altaffilmark{1}}

\author{Christopher W. Churchill}
\affil{Department of Astronomy and Astrophysics \\
       The Pennsylvania State University \\
       University Park, PA 16802 \\
       {\it cwc@astro.psu.edu}} 

\vglue 0.1in
\author{Steven S. Vogt}
\affil{UCO/Lick Observatories \\
       Board of Studies in Astronomy and Astrophysics \\
       University of California, Santa Cruz, CA 96054 \\
       {\it vogt@ucolick.org}}

\vglue 0.1in
\and
\vglue -0.1in
\author{Jane C. Charlton\altaffilmark{2}}
\affil{Department of Astronomy and Astrophysics \\
       The Pennsylvania State University \\
       University Park, PA 16802 \\
       {\it charlton@astro.psu.edu}}

\altaffiltext{1}{Based  in  part   on  observations  obtained  at  the
W.~M. Keck Observatory, which  is operated as a scientific partnership
among Caltech, the University of California, and NASA. The Observatory
was made possible by the  generous financial support of the W.~M. Keck
Foundation.}
\altaffiltext{2}{Center for Gravitational Physics and Geometry}

\begin{abstract}

We present a detailed and statistical analysis of the column densities
and Doppler $b$ parameters of {\MgII} absorbing clouds at redshifts
$0.4 \leq z \leq 1.2$.  We draw upon the HIRES/Keck data ($\Delta v
\simeq 6.6$~{\kms}) and Voigt profile (VP) fitting results presented
by Churchill \& Vogt (Paper I).  The sample is comprised of 175 clouds
from 23 systems along 18 quasar lines of sight.  In order to better
understand whether the inferred physical conditions in the
absorbing clouds could be ``false'' conditions, which can arise due to the
non--uniqueness inherent in parameterizing complex absorption
profiles, we performed extensive simulations of the VP analyses
presented in this paper.  In brief, we find: (1) The {\FeII} and
{\MgII} column densities are correlated at the 9~$\sigma$ level.
There is a 5~$\sigma$ anti--correlation between the {\MgI}/{\MgII}
column density ratio and the {\MgII} column density.  (2) Power--law
fits to the column density distributions for {\MgII}, {\FeII}, and
{\MgI} yielded power--law slopes of $\simeq -1.6, -1.7$, and $-2.0$,
respectively.  (3) The observed peaks of the Doppler parameter
distributions were $\sim 5$~{\kms} for {\MgII} and {\FeII} and $\sim
7$~{\kms} for {\MgI}.  The clouds are consistent with being thermally
broadened, with temperatures in the 30--40,000~K range.  (4) A
two--component Gaussian model to the velocity two--point correlation
function (TPCF) yielded velocity dispersions of $54$~{\kms} and
$166$~{\kms}.  The narrow component has roughly twice the amplitude of
the broader component.  The width and amplitude of the broader
component decreases as equivalent width increases.  (5) From
photoionization models we find that the column density ratios are most
consistent with being photoionized by the ultraviolet extragalactic
ionizing background, as opposed to stellar radiation.  Based upon the
{\MgI} to {\MgII} column density ratios, it appears that at least
two--phase ionization models are required to explain the data.

\end{abstract}


\section{Introduction}
\label{sec:intro}

In \citet[][hereafter Paper I]{cv01}, we presented 23 {\MgII}
absorption systems observed at high--resolution in high
signal--to--noise quasar spectra.  The analysis presented in Paper~I
was primarily based upon direct measurements of the data.  In this
paper we concentrate on measurements obtained using Voigt profile
fitting of the data.

Voigt profile analysis provides a well defined, if simplistic,
parameterization of complex absorption lines.  From the beginning of
the last decade \citep[e.g.,][]{carswell91}, Voigt profile (VP)
analysis has been a useful technique for statistically documenting the
physical conditions of absorbing gas in quasar spectra.  VP analysis
yields the number of ``clouds'' in a given absorption complex, their
Doppler parameters, and their relative line--of--sight velocities.
Since the atomic physics is incorporated, the line--of--sight
integrated column densities and the thermal and/or turbulent
properties of the gas clouds can be measured.

The reported statistics of both the high redshift {\Lya} forest
\citep{hu_lya95,lu_lya96} and {\CIV} absorbers \citep{rauch-civ} have
been based upon VP decomposition of high resolution ($R\simeq 45,000$)
and signal--to--noise ratio ($S/N \simeq 30-80$) spectra.  A primary
assumption of VP analysis is that complex absorption profiles can be
decomposed into a product of individual components, each of which are
treated as spatially isolated, isothermal clouds intercepting the
incoming flux at a unique line--of--sight velocity.  However,
numerical simulations of the {\Lya} forest and of {\CIV} systems at
high redshift undermine this physical picture because the absorbing
systems merge continuously into a smoothly fluctuating background,
often are comprised of a range of temperatures, and are usually
broadened by coherent velocity flows \citep{haehnelt96,rad97}.  This
is not altogether surprising for higher redshifts, given that {\Lya}
and {\CIV} absorption likely arise in large, coherent structures that
are still coupled to the Hubble flow \citep[e.g.,][]{rhs97}.

Unblended absorption lines arising in lower ionization
gas, as traced by the {\MgIIdblt} doublet, are often relatively narrow
and suggestive of quiescent single ``clouds''
\citep{weakI,q1206,cv01}.  As such, the low ionization gas more
directly associated with the components of galaxies may be a better
match to the assumptions underlying VP analysis.  However, {\MgII}
systems also show a wide variation in kinematic complexity, including
significant blending of the profiles arising from the discrete clouds
\citep{pb90,cv01}.  It is not clear how this particular type of
blending could systematically skew the statistical results yielded by
VP analysis.  In addition, an unavoidable consequence of VP analysis
is that the results are sensitive to both the resolution and
signal--to--noise ratio of the spectra \citep{thesis,rauch1422}.

Even with the many caveats mentioned above, the potential for VP
analyses to answer many questions regarding the physical nature of
the absorbing gas remains widely recognized.  What is the average number of
clouds per system?  What is the distribution of column densities and
column density ratios?  How does the cloud--cloud velocity clustering
depend upon equivalent width?  Are the line broadening mechanisms
thermal, or dominated by turbulent motions in the gas?  With the aid
of state of the art photoionization models we can ask: What are the
chemical and ionization conditions of the studied ionic species?  What
can be inferred about the nature of the ionizing radiation?  If there
is a stellar contribution, what limits can be placed on the numbers of
stars and their distances from the gas clouds?

For {\MgII} systems, answers to some of these questions have been
investigated using data with a resolution of $~30$~{\kms}
\citep{pb90,bergeron94}.  Since VP analysis is strongly dependent upon
spectral resolution, it is a logical step to re--examine the
statistical distributions previously inferred for {\MgII} systems with
higher resolution data.  In this paper we present statistical results
from a VP analysis of an ensemble of 23 {\MgII}--selected absorbers
obtained with the HIRES instrument \citep{vogt94} on the Keck~I
telescope.  The resolution of these data is $\simeq 6$~{\kms}.  The
systems and the VP results were previously presented in Paper~I.

In \S~\ref{sec:data}, we briefly review the data.  In
\S~\ref{sec:vpanal}, we describe the VP analysis of the data, our VP
decomposition methodology, and simulations of the analysis for
purposes of diagnosing whether or not a given statistical result is
induced, in part or in full, by the VP technique itself.  In
\S~\ref{sec:column}, we present the distribution of VP column
densities, Doppler parameters, and velocities.  We also provide some
analysis of these distributions.  Photoionization models of the VP
components, or clouds, are discussed in \S~\ref{sec:models}.  We
include an exploration of the possible relative contribution of
stellar flux to that of a standard ultraviolet background.  We outline
the main results of this paper in \S~\ref{sec:conclude}.

\section{The Data}
\label{sec:data}

The acquisition, calibration, and analysis of the data has been
described in \citet{thesis} and in Paper~I.  However, we provide
a brief description here.  We emphasize that all systems are selected
by the presence of {\MgIIdblt} absorption.

The   quasar  spectra   were  obtained   with  the   HIRES  instrument
\citep{vogt94} on  the Keck~I telescope.  The  spectral resolution is
$R=45,000$.   The signal--to--noise  ratio in  the continuum  near the
{\MgII} absorption  profiles is  typically $\sim 30$.   A total  of 23
systems comprise  the sample of  absorbers.  The sample  includes only
those  systems  with  {\MgII}  $\lambda 2796$  rest--frame  equivalent
width,  $W_{r}(2796)$,  greater  than  $0.3$~{\AA}.  The  ensemble  of
systems is considered to be  representative of a ``fair'', or unbiased
sample,  in that  the  distribution of  {\MgII}  equivalent widths  is
consistent  with   that  of  unbiased   surveys
\citep[e.g.,][]{ltw87,ss92,weakI}. 

The sample has been divided into four subsamples by their equivalent
widths.  Sample B has $W_{r} = (0.3,0.6]$~{\AA}, sample C has $W_{r} =
(0.6,1.0]$~{\AA}, sample D has $W_{r} \geq 0.6$~{\AA}, and sample E
has $W_{r} \geq 1.0$~{\AA}.  We have used the notation ``('' to denote
inclusive and ``]'' to denote exclusive.  These samples are historically
motivated \citep[see][]{ss92}.  In Table~\ref{tab:samples} we present
the systems by quasar, absorber redshift, and sample membership.  Also
listed is a system identification, or ID.

\begin{figure*}[p]
\plotone{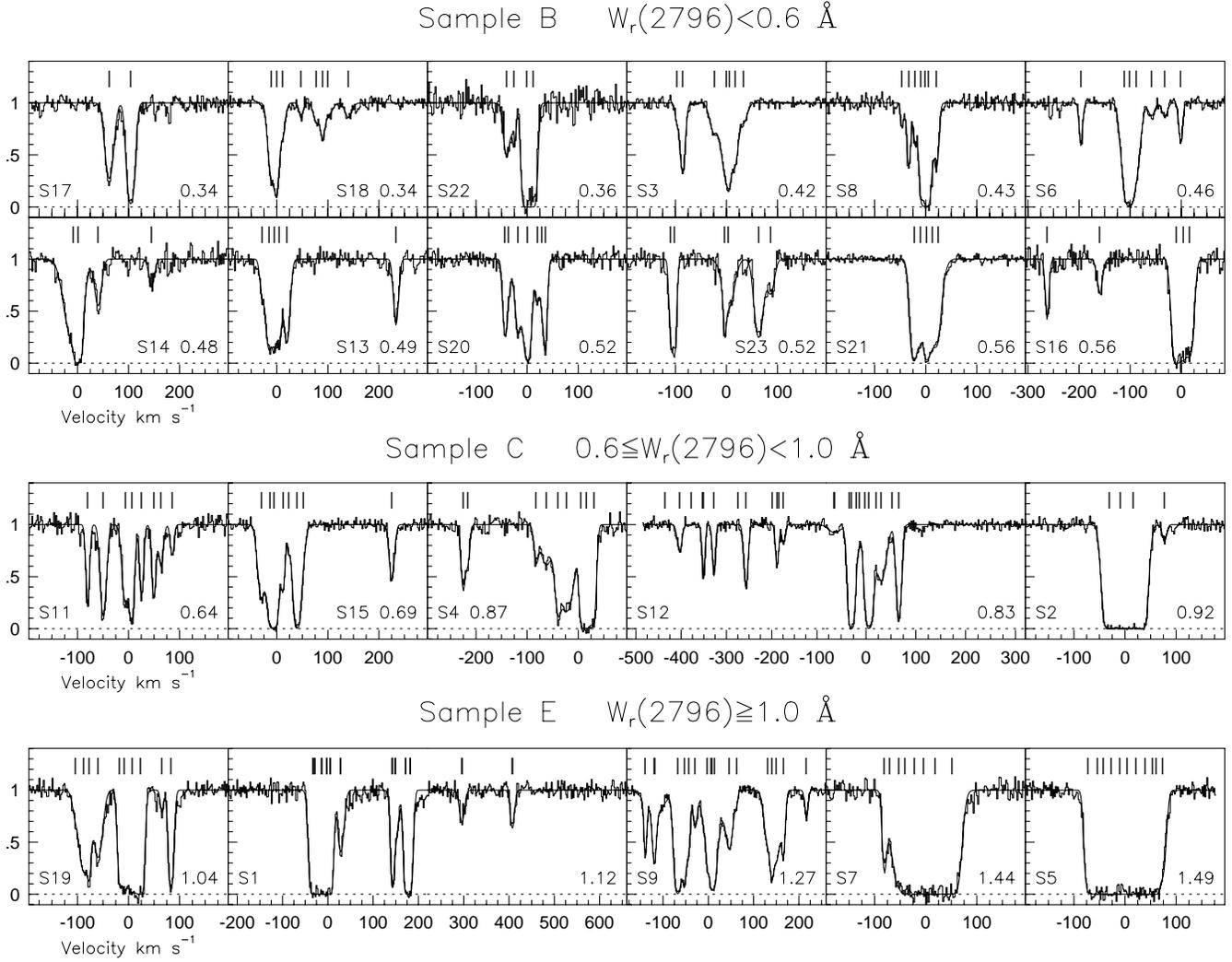}
\figurenum{1}
\caption
{Normalized HIRES/Keck spectra of the {\MgII} $\lambda 2796$
absorption profiles presented in the system rest--frame velocity.  The
solid curves through the data (histogram) are the model spectra from
Voigt profile (VP) fitting using the code MINFIT.  The ticks above
each continuum give the VP velocity centroids.  The systems are
presented by sample membership in order of increasing rest--frame
equivalent width.  Each system identification (see
Table~\ref{tab:samples}) and the equivalent width are given in the
bottom part of the panels.
\label{fig:data}}
\end{figure*}

The {\MgII} $\lambda 2796$ profiles are presented in
Figure~\ref{fig:data}.  The spectra are normalized and converted to
rest--frame velocity.  The systems are grouped by sample membership
and are ordered by increasing $W_{r}$.  The system ID and equivalent
width are given in each panel.  Each system has been modeled using
Voigt profile (VP) analysis, as described in Paper~I.  The solid
curves through the data are the VP model spectra and the ticks above
the spectra are the VP centroids.

\section{Voigt Profile Analysis}
\label{sec:vpanal}

We used  the Voigt profile  fitting program MINFIT  \citep{thesis} to
obtain a  least--squares fit (LSF),  $\chi ^{2}$ minimized  model of
each absorption line profile.  MINFIT sets up M nonlinear functions in
N  parameters   and  drives  the  NETLIB  slatec   LSF  routine  DNLS1
\citep{more78,hiebert80}.   The   N   parameters  are   the
redshifts, column  densities, and Doppler  $b$ parameters for  each VP
component.  The MINFIT program statistically minimizes the number of VP
components used to model a given system.  Numerical convergence is set
to  the machine  precision  of  a Sun  Ultra~1.   Further details  are
provided in \citet{thesis}.

As discussed in Paper~I, we concentrate our analysis on the
{\MgIIdblt} doublet, the {\MgI} $\lambda 2852$ transition, and the
{\FeII} $\lambda 2344$ $2374$, $2582$, $2587$, and $2600$ multiplet.
The atomic data used for the VP analysis are given in Table~2 of
Paper~I.  The column densities, $b$ parameters, and cloud velocities
output by MINFIT are given in Table~7 of Paper~I and illustrated for
all transitions in Figure~2 of Paper~I.

\begin{figure*}[p]
\plotone{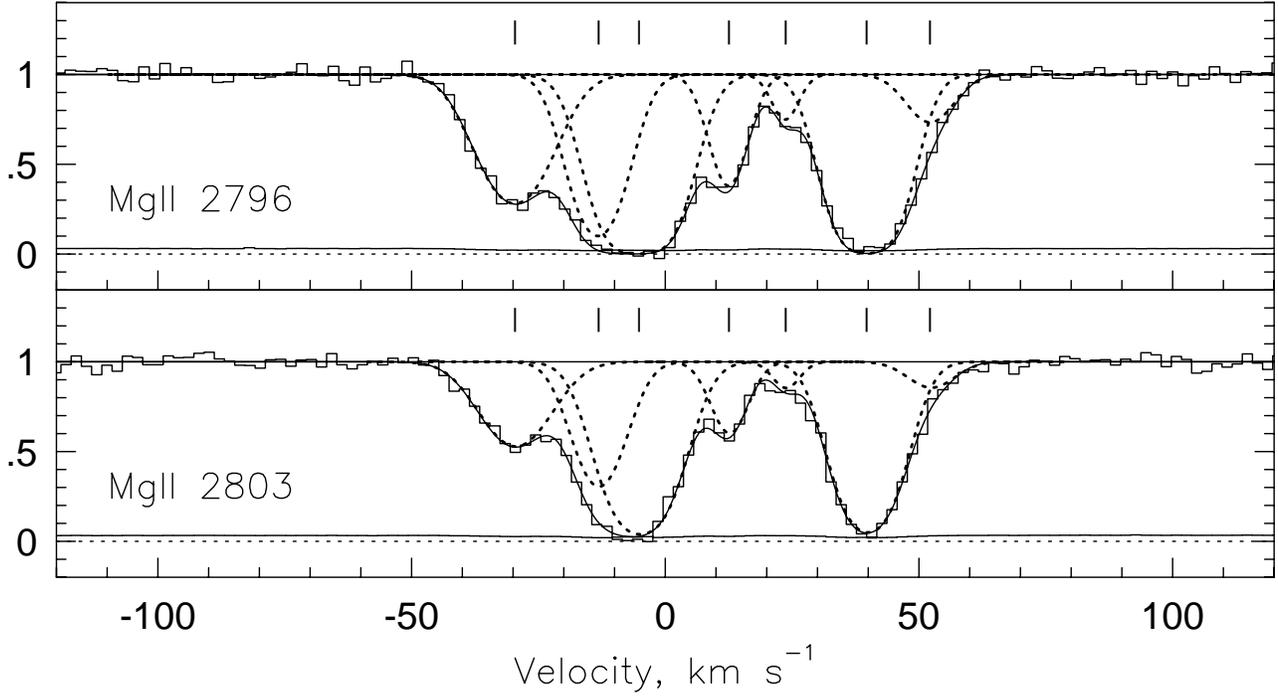}
\figurenum{2}
\caption
{An illustration of the Voigt profile (VP) modeling of the {\MgIIdblt}
absorption profiles for the $z=0.7730$ system in the spectrum of
PG~$1248+401$ (ID S15).  The solid curves through the data
(histogram) are the model spectra from VP fitting using the code
MINFIT.  The ticks above each continuum give the VP velocity centroids.
The individual VP components are shown as dotted curves.  The ``high
velocity'' cloud in this systems at $v=+225$~{\kms} is not shown.
\label{fig:vpfit}}
\end{figure*}

The resulting VP model spectra for the {\MgII} $\lambda 2796$
transitions are shown in Figure~\ref{fig:data} as solid curves through
the data.  Ticks above the spectra give the number of VP components
and their rest--frame velocities.  In Figure~\ref{fig:vpfit}, we show
a detail of the fit for both members of the {\MgII} doublet for system
S15.  For this system, seven individual VP components (dotted curves)
successfully model the complex profile shape.  For the illustrated velocity
region, the reduced chi--square was $\chi ^{2}_{\nu} = 1.3$ for $418$
degrees of freedom (including one {\MgI} and five {\FeII} transitions
each with 50--60 pixels). 

Since  VP  analysis yields  a  non--unique  interpretation of  complex
absorption  lines, it  is important  to outline  key specifics  of the
fitting procedure and calibration of the results if the analysis is to
be  considered objective.  In  the following,  we briefly  outline the
fitting methodology  applied in this  work, with focus on  specifics of
the  operations  performed  by  the  MINFIT code.   We  also  describe
simulations of  VP analysis that  we use to calibrate  our statistical
results.

\subsection{Profile Fitting Methodology}

\subsubsection{The Number of VP Components}

MINFIT  employs significance testing  in order  to return  the minimum
number  of VP  components.  Two  user--defined,  run--time parameters,
$f_{err}$ and $f_{c}$, provide the maximum fractional error allowed in
a VP  component and the  confidence level at  which a VP  component is
considered statistically significant, respectively.  These parameters
are  adjustable,  which adds  a  subjective  element  to an  otherwise
objective, well  defined algorithm.  Typical values  for $f_{err}$ and
$f_{c}$ are 1.5 and 0.97, respectively.

Once  a  LSF solution  is  obtained,  MINFIT  computes errors  in  the
standard fashion using the diagonal elements of the covariance matrix,
which are  obtained from the  inversion of the curvature  matrix.  The
elements  of  the  curvature  matrix  are  computed  as  described  in
\citet{thesis}.

The fractional error of each VP component is the quadrature sum of the
fractional errors in the three VP parameters, $N$, $b$, and $z$.  When
the fractional  error of a component exceeds  $f_{err}$, the component
is  dropped.  The  LSF and  error computations  are iterated  until no
components have fractional errors exceeding $f_{err}$.
  
At  this  stage  the  significance  level is  determined  for  the  VP
component with the largest  fractional error.  The target VP component
is  dropped   and  a   new  LSF  solution   is  obtained.    Then,  an
$F$--distribution test is performed between the two LSF solutions.  If
the VP component was significant with a confidence level of $f_{c}$ or
better,  than  the  original  LSF  model is  restored.   If  not,  the
statistical significance  is determined for the VP  component with the
largest fractional error  in the new model.  This  process is repeated
until all VP components are  significant at a minimum confidence level
of $f_{c}$.

\vglue 0.5in
\subsubsection{Redshifts}

It was assumed that the redshifts, $z$, and therefore rest--frame
velocities, of each VP component are aligned for all ions.  VP
components were defined by the presence of {\MgII} absorption features
even if there were no {\FeII} or {\MgI} absorption features at the
same velocity.

We note that the ionization potential of {\MgI} is below the hydrogen
edge.  The adopted methodology of enforcing an aligned velocity for
all ions may not be sound if the gas has an ionization structure that
correlates with line of sight velocity.  That is, {\MgI} could, in
principle, arise from a physically and/or kinematically distinct part
of the clouds in comparison to {\MgII} and {\FeII}.

\subsubsection{Column Densities}

For each VP component, the column densities, $N$, of different ions
are determined independent of one another.  When multiple transitions
of a given ion are available, then the increased number of constraints
yield a more robust measurement.  For {\MgI} there is only the
singlet.  For {\MgII} there is always the resonance doublet and for
{\FeII} there are often three to five transitions of the multiplet
available.  As such, {\MgI} is the least constrained ion.  

If neither {\MgI} nor {\FeII} had detectable absorption at the same
velocity as a {\MgII} component, then an upper limit on the {\MgI} or
{\FeII} column density is quoted for the component.  This limit is
taken from the 3~$\sigma$ equivalent width threshold for an
unresolved absorption line at the expected location of the component
\citep[see][]{archiveI}.

\subsubsection{Doppler Parameters}

The line broadening, or $b$ parameter, of a VP component can be
written as the convolution of a thermal component and a turbulent
component.  In order to preserve the Voigt profile formalism, one is
required, perhaps quite unrealistically, to parameterize the
turbulence with a Gaussian distribution, yielding $b^{2} =
b_{therm}^{2} + b_{turb}^{2}$.  Whereas $b_{therm}$ scales with the
inverse square root of the ion mass, $b_{turb}$ is independent of ion
mass and will have the same magnitude for all ions.

There are three possible treatments of the $b$ parameters.  In each VP
component, one can: (1) measure the $b$ parameter independently for
each ion with no assumptions about thermal scaling or level of
turbulence, (2) force the $b$ parameter to be equal for all ions,
which holds if the turbulent component dominated, or (3) force the
turbulent component to be some fraction of the measured $b$ parameter
(zero for pure thermal).  We assume the first approach, so that we can
fully explore the line broadening mechanism.

\subsection{Calibrating Our VP Analysis}

The only way to calibrate VP analysis is to quantify the degree to
which the fitted parameters reflect the ``true'' cloud properties in
cases where they are known {\it a priori}.  This requires simulations
for which the VP parameters obtained from analysis of synthetic
spectra can be directly compared to the VP parameters used to generate
these spectra.

We have run a series of comprehensive simulations of {\MgII} absorbers
for varying degrees of line blending (i.e.\ complexity of absorption
profiles) and for a range of spectral signal--to noise ratios (i.e.\
equivalent width detection threshold) commensurate with the data.
These simulations have been described in detail in \citet{thesis}, and
serve to quantify any possible bias, systematic, or random errors in
the VP results to the data.  For each analysis using VP fits to the
observational data, we performed an identical analysis using VP fits
to the synthetic spectra.  In this way, the systematics or
significance of any observational statistical result can be quantified
and any systematic skew in the functional form of a distribution can
be estimated.

The simulations required input column density, $b$ parameter, and
velocity clustering distributions.  We used the data to guide our
``best'' choices of these distribution functions by comparing (1) the
distribution of pixel flux decrements, (2) the distribution of
equivalent widths of the kinematic subsystems\footnote{Kinematic
subsystems were defined in Paper~I.  They are the individual
absorption features within $\pm 500$~{\kms} of the {\MgII} absorption
centroid that are separated by a minimum of three pixels (a resolution
element) of continuum flux.}, (3) the distribution of velocity widths
of the kinematic subsystems, and (4) the observed VP component column
density, $b$ parameter, and velocity distributions.

The adopted distributions for the simulations were obtained by
constructing a set of 500 simulated absorption systems and performing
VP analysis on these spectra.  We then performed statistical tests
comparing the simulation distributions and the observed distributions
to eliminate those which were not statistically consistent with data
\citep[also see][]{kinematicpaper}.  For the {\MgII} column densities,
we chose a power law distribution, $f(N) \propto N^{-\delta}$ with
$\delta = 1.6$.  The $b$ parameters distribution was drawn from a
Gaussian with centroid $4$~{\kms}, a width of $\sigma = 1.5$~{\kms}
and a lower cut--off of $1.5$~{\kms}.  The lower cut off was required
to match the small $b$ portion of the observed distribution.  For the
VP component velocity clustering, we chose a Gaussian distribution
with $\sigma = 45$~{\kms}.

We  also  simulated  the   {\FeII}  multiplet.   We  enforced  thermal
broadening for all  VP components, i.e.\ the $b$  parameters scaled as
$b({\FeII}) = \{ m({\rm Mg})/m({\rm Fe}) \} ^{1/2}b({\MgII})$, where $m$
is the  atomic mass.  We  also assumed a linear relationship
between $\log N({\FeII})$ and $\log N({\MgII})$ for all components.
This relation, $\log  N({\FeII}) = \log N({\MgII}) - 0.3$,
is  based upon  preliminary  VP  fitting to  the  sample presented  in
\citet{thesis}.

Similar simulations for VP analysis calibration have been undertaken
by \citet{lu_lya96} and by \citet{hu_lya95} for the {\Lya} forest.
The significant difference between the simulation performed here and
those for the {\Lya} forest are that (1) the hydrogen
absorption--selected clouds were assumed to be distributed
cosmologically, whereas the {\MgII} absorbing clouds are clustered in
kinematic systems, and (2) the VP analysis of the {\MgII} systems
included simultaneous fitting to the {\MgII} doublet and {\FeII}
multiplet.


\section{Results of Voigt Profile Analysis}
\label{sec:column}

\subsection{Number of VP Components (i.e.\ Clouds)}

\begin{figure*}[p]
\plotone{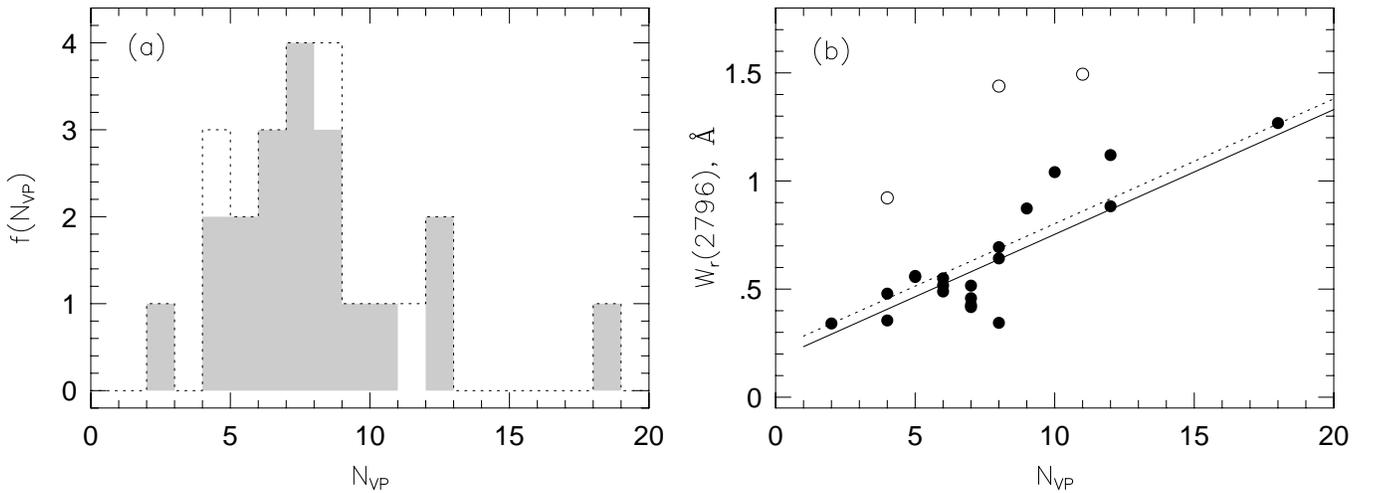}
\figurenum{3}
\caption
{(a) The distribution of VP components.  The solid--line histogram
represents the entire sample, whereas the shaded histogram represents
the same sample but with the DLA and {\HI}--rich systems removed.
The average number of clouds per absorber is 7.7 for both
distributions.  --- (b) The rest--frame {\MgII} $\lambda 2796$
equivalent width versus the number of VP components.  Two least
squares fits are shown.  The dotted line includes the full sample; the
solid line is for the the same sample but with DLA and {\HI}-rich
systems removed.  See text for details.
\label{fig:nclouds}}
\end{figure*}

In Figure~\ref{fig:nclouds}$a$ we show the distribution of the number
of VP components (i.e. clouds).  The dotted--line histogram represents
the full sample of 23 systems.  The shaded histogram represents a
subset of the full sample in which the very large $N({\HI})$ systems
(S2, S5, and S7) have been removed.  These damped {\Lya} absorbers
(DLAs) and ``{\HI}--rich'' systems \citep{archiveII} have been removed
in light of their characteristic black--bottomed {\MgII} absorption
profile morphologies; it is difficult to constrain the number of
clouds in these systems (they are systematically underestimated unless
there are good constraints from weak {\FeII} transitions).  For ease
of discussion, we will refer to this subset of our 23 systems as the
``non--DLA'' subset.

For both the full sample and the non--DLA subset, there is an average
of 7.7 clouds per absorber.  The mode of the binned
distribution is eight clouds.  In the simulations of complex, blended
profiles, $\sim 30$\% of the inputted VP components were not
recovered.  These simulations also revealed that the percentage of
unrecovered components decreases with decreasing signal--to--noise
ratio, but that the width of the recovered distribution becomes larger
as the signal--to--noise ratio increases.  To the extent that VP
components (such as those in our models) give rise to the observed
absorption profiles, the distribution plotted in
Figure~\ref{fig:nclouds}$a$ would require a correction such that the
average and mode is elevated to 10--11 clouds per system.

In Figure~\ref{fig:nclouds}$b$, we show the rest--frame equivalent
width of {\MgII} $\lambda 2796$ versus the number of VP components,
$N_{cl}$.  The unfilled data points are the DLA and {\HI}--rich
systems.  As has been reported previously \citep{pb90,csv96,thesis},
there is a very strong correlation between the equivalent width and
the number of VP components.  A linear LSF, including errors in the
equivalent widths, was performed on the data presented in
Figure~\ref{fig:nclouds}$b$.  The dotted line is the fit to the full
sample, which yielded a slope of $0.058\pm0.004$~{\AA}/cloud and an
equivalent width of $0.28\pm0.03$~{\AA} for $N_{cl} = 1$ (the
intercept at $N_{cl}=0$ has no physical meaning).

A fit to the non--DLA subset yielded a statistically identical slope
with a slightly smaller value of $0.23\pm0.03$~{\AA} for $N_{cl}=1$.
This fit is shown as the solid line on Figure~\ref{fig:nclouds}$b$.
Thus, at least in this sample of 23 systems, the three DLA and
{\HI}--rich systems serve only to slightly increase the intercept to
the fit and do not skew the slope.  We also examined whether the data
point at $N_{cl}=18$ was dominating the fit results by removing it and
refitting to the non--DLA subset of data; the resulting fit was
statistically identical to the fit to the full sample.

The slopes of the linear relationship presented here are slightly
smaller than both the slopes of $0.070\pm0.004$~{\AA}/cloud reported
by \citet{csv96} for a smaller sample of 15 systems and $0.076\pm
0.004$~{\AA}/cloud reported by \citet{thesis} for a sample of 36
systems.  This difference is most likely due to the fact that in this
work we did not enforce the fit to pass through the origin, whereas
the fit was forced through the origin in the previous work.  We stress
that the slope is strongly dependent upon the spectral resolution of
the data.  For example, \citet{pb90} obtained a much steeper slope of
$0.35$~{\AA}/cloud for a large sample observed at a resolution of
$\simeq 30$~{\kms}.

\vglue 0.5in
\subsection{Column Densities}

\begin{figure*}[p]
\plotone{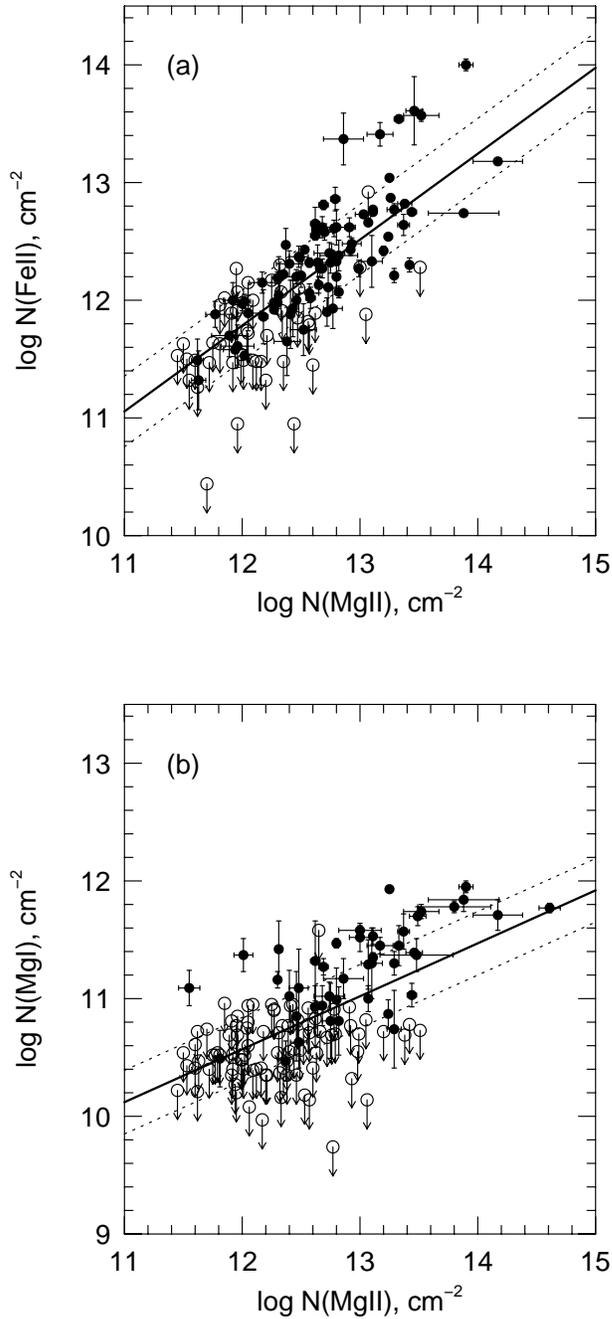}
\figurenum{4}
\caption
{(a) The logarithmic VP column densities (in atoms~{\cmsq}) for {\FeII} as
a function of {\MgII}.  Upper limits for {\FeII} are shown as open
data points.  The best--fit slope, shown as a solid line, is
$0.73\pm0.06$.  The reduced $\chi ^{2}$ of $30$ provides a measure of
the scatter about this relationship.  The standard deviation of the
fit is shown by the two dotted lines.  --- (b)
The logarithmic VP column densities (in atoms~{\cmsq}) for {\MgI} as
a function of {\MgII}.  Upper limits for {\MgI} are shown as open data
points.  The best--fit slope, shown as a solid line, is $0.45\pm0.05$.
The reduced $\chi ^{2}$ of $27$ provides a measure of the scatter
about this relationship.  The standard deviation of the fit is shown
by the two dotted lines.
\label{fig:mg2+fe2+mg1}}
\end{figure*}

In Figure~\ref{fig:mg2+fe2+mg1}, we  present {\FeII} and {\MgI} column
densities vs.\  {\MgII} column  densities.  Only components  for which
the errors are less than 0.7 dex are plotted.  Upper limits
on the  {\FeII} and {\MgI} column  densities are shown  as open points
with downward pointing  arrows and measured values are  shown as solid
points with $1~\sigma$ error bars.

$N(\hbox{\FeII})$  is strongly  correlated with  $N(\hbox{\MgII})$.  A
Spearman--Kendall test  yields a $9~\sigma$  significance.  Assuming a
linear relationship of the form $\log N({\rm X}) = a \log N({\MgII}) +
b$,  we  determined  the  best  fit parameters  using  a  $\chi  ^{2}$
minimization  that allowed  for  errors in  the  {\FeII}, {\MgI},  and
{\MgII}  column   densities.   The  best  fit  parameters   are  $a  =
0.73\pm0.06$ and  $b = 3.0\pm 0.8$  for {\FeII} and  $a = 0.45\pm0.05$
and $b = 5.2\pm 0.6$ for {\MgI}.  Limits have been included in  the
fitting.  

The reduced $\chi ^{2}$ is $30$ for {\FeII} and is $27$ for {\MgI}.
Fits to the {\FeII} vs.\ {\MgII} and {\MgI} vs.\ {\MgII} column
densities obtained from the simulations yielded $\chi ^{2} \simeq 10$.
Thus, the variance in the data is roughly a factor of three greater
than the variance due to statistical scatter inherent in the VP
fitting.

\begin{figure*}[p]
\plotone{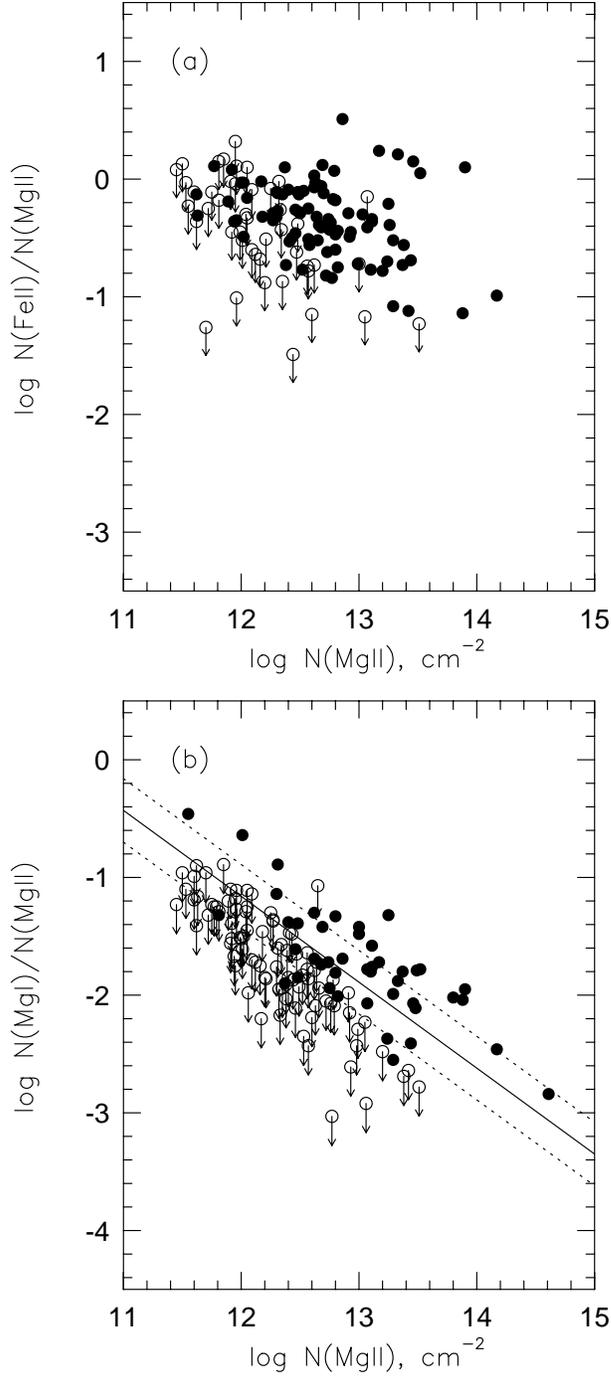} 
\figurenum{5}
\caption
{(a) The logarithmic VP column density ratio of {\FeII} to {\MgII} as
a function of {\MgII}.  The error bars are not shown for presentation
purposes; errors and data point types can be viewed in
Figure~\ref{fig:mg2+fe2+mg1}$a$. --- (b) The logarithmic VP column
density ratio of {\MgI} to {\MgII} as a function of {\MgII}.  The
best--fit slope, shown as a solid line, is $-0.73\pm0.06$.  The
standard deviation of the fit is shown by the two dotted lines. The
error bars are not shown for presentation purposes; errors and data
point types can be viewed in Figure~\ref{fig:mg2+fe2+mg1}$b$.
\label{fig:mg2-fe2-mg1}}
\end{figure*}

In Figure~\ref{fig:mg2-fe2-mg1}, we present {\FeII} to {\MgII} and
{\MgI} to {\MgII} column density ratios vs.\ {\MgII} column densities.
The data presented are the same VP components shown in
Figures~\ref{fig:mg2+fe2+mg1}$a$ and \ref{fig:mg2+fe2+mg1}$b$.  Upper
limits on the ratios are shown as open points with downward pointing
arrows and measured values are shown as solid points.  The error bars
are suppressed, but they are effectively identical to those presented
in Figure~\ref{fig:mg2+fe2+mg1}.

The distribution of $N({\FeII})/N({\MgII})$ visually appears to be a
slightly decreasing function of $N({\MgII})$, but is statistically
consistent with no dependence upon $N({\MgII})$.  This would indicate
that $N({\FeII})/N({\MgII})$ is fairly independent of or only slightly
decreasing with increasing $N({\MgII})$.  The greater than 1~dex
spread in the ratio would indicate a range of [$\alpha$/Fe] abundance
patterns, dust depletion factors, ionization conditions, or a
combination of all three.

There is a significant ($5~\sigma$) anti--correlation between
$N({\MgI})/N({\MgII})$ and $N({\MgII})$ over a 2.5 order of magnitude
range of the ratio.  For the upper limits, taken by themselves, this
trend is an artifact of the limiting detectable $N({\MgI})$, which is
effectively a constant independent of $N({\MgII})$.  However, for the
measured ratios, the anti--correlation is real and cannot easily be
explained as an artifact of the VP analysis of the data.  A linear LSF
to the detections only (filled data points of lower panel in
Figure~\ref{fig:mg2-fe2-mg1}) yielded a slope of $a= -0.73\pm 0.06$
and intercept $b=7.6\pm0.3$.  The linear fit and its uncertainty are
superimposed on the data.

Further discussion of the column density ratios is presented in
\S~\ref{sec:models} in the context of photoionization models.

\subsection{Column Density Distribution}
\label{sec:maxlike}

VP analysis of simulated spectra with comparable signal--to--noise
ratios revealed that the input column density distribution is recovered
within the $1~\sigma$ errors when an appropriate lower--limit,
cut--off column density is applied to a maximum likelihood fit.
This cut off is the column density below which the detection
completeness drops rapidly.

The simulations further revealed that the completeness level is
sensitive to the kinematic clustering of components due to line
blending.  In regions of significant blending the 90\% completeness
levels are $\log N({\MgII}) = 12.4$~{\cmsq} and $\log N({\FeII}) =
12.2$~{\cmsq} (we did not examine blending effects for {\MgI}).  The
90\% completeness levels for unblended lines are $\log N({\MgII}) =
11.6$~{\cmsq}, $\log N({\FeII}) = 11.8$~{\cmsq}, and $\log N({\MgI}) =
11.4$~{\cmsq}.  Thus, there is a column density range of ``partial
completeness'', a transition from incompleteness due to line blending
to incompleteness due to finite signal-to--noise ratio.

\begin{figure*}[p]
\plotone{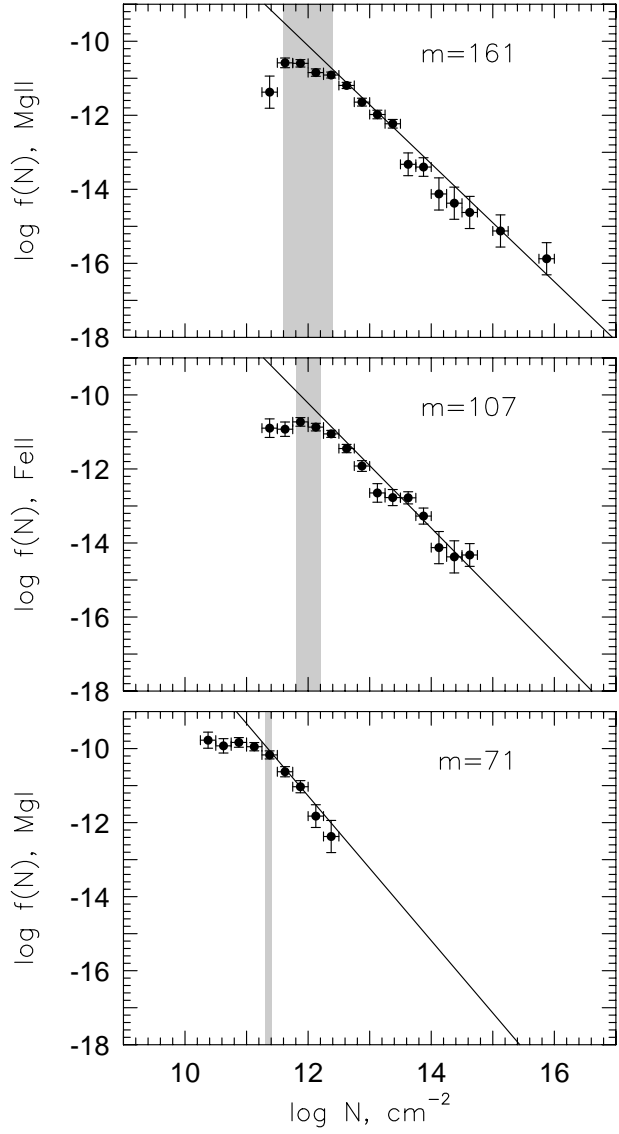}
\figurenum{6}
\caption
{The logarithmic VP column density distribution functions for {\MgII}
(top), {\FeII} (center), and {\MgI} (bottom).  The maximum likelihood
results (unbinned data) for the relation $f(N) \propto N^{-\delta}$
are $\delta = 1.59\pm0.05$, $1.69\pm0.07$, and $2.02\pm0.03$ for
{\MgII}, {\FeII}, and {\MgI}, respectively.  The number of clouds,
$m$, in each sample is written in each panel.  The shaded areas
show the column density ranges where line blending leads to ``partial
completeness'' (see text).
\label{fig:ndist}}
\end{figure*}

In Figure~\ref{fig:ndist}, the column density distribution functions
are shown for {\MgII}, {\FeII}, and {\MgI}.  The data are binned for
purpose of presentation.  The shaded regions show the column density
ranges of partial completeness.  The data appear to follow a
power--law distribution, with the turnover at small column densities
due to incompleteness.  At HIRES/Keck resolution, an unresolved
{\MgII} line saturates at $\log N \simeq 13.5$~{\cmsq} and this
explains the increased scatter in the number density of {\MgII} VP
components with column densities greater than this value \citep[see
Figure 4.3 of][]{thesis}.

For the column density distribution, we assumed a power law
\begin{equation}
 f(N) = CN^{-\delta} ,
\label{eq:ndistfunc}
\end{equation}
where $f(N)$ is the number of clouds with column density $N$ per unit
column density.  The total number of clouds, $m$, 
normalizes $f(N)$ over the observed data according to
\begin{equation}
  \int _{N_{\rm min}} ^{N_{\rm max}} f(\hat{N})d\hat{N}
  = m .  
\end{equation}
We used the maximum likelihood method \citep[e.g.,][]{tytler87,ltw87}
to obtain the power--law slope and normalization.  The method is
performed on the unbinned data.  In Figure~\ref{fig:ndist}, the solid
lines through the data are the maximum likelihood solutions, which are
\begin{equation}
{\MgII}: \quad \delta = 1.59 \pm 0.05 \quad C = (9.85 \pm 0.16) \times
10^{8} ,
\end{equation}
\begin{equation}
{\FeII}: \quad \delta = 1.69 \pm 0.07 \quad C = (1.01 \pm 0.22) \times
10^{10} ,
\end{equation}
\begin{equation}
{\MgI}: \quad \delta = 2.02 \pm 0.28 \quad C= (6.21 \pm 0.79) \times
10^{12} .
\end{equation}
For this analysis, we used the cut--off column densities appropriate
for blended lines, which yielded $m=161$, $107$, and $71$ for {\MgII},
{\FeII}, and {\MgI}, respectively.

Using the parameterization given in Equation~\ref{eq:ndistfunc},
\citet{pb90} found a $\delta = 1.0\pm0.1$ for {\MgII} over the range
$12.0 \leq \log N(\hbox{\MgII}) \leq 14.3$~{\cmsq}.  The resolution of
their data was $\sim 30$~{\kms}, and it is likely that their best fit
slope is affected by unresolved blending.  The expected trend is that
the number of large column density components would decrease, being
redistributed into a greater number of smaller column density
components as the resolution is increased.  This would result in a
steeper slope, as we have found in this work.

The slopes of the {\MgII} and {\FeII} column density distributions are
very similar, and roughly consistent with one another.  However, the
slope of {\MgI} distribution is significantly steeper.  These findings
are consistent with the distribution of column density ratios
presented in Figure~\ref{fig:mg2-fe2-mg1}.  There is no correlation of
$N({\FeII})/N({\MgII})$ with $N({\MgII})$, yet there is a significant
anti--correlation of $N({\MgI})/N({\MgII})$ with $N({\MgII})$.

\subsection{Doppler Parameter Distribution}
\label{sec:doppler} 
\label{sec:bdist}

\begin{figure*}[p]
\plotone{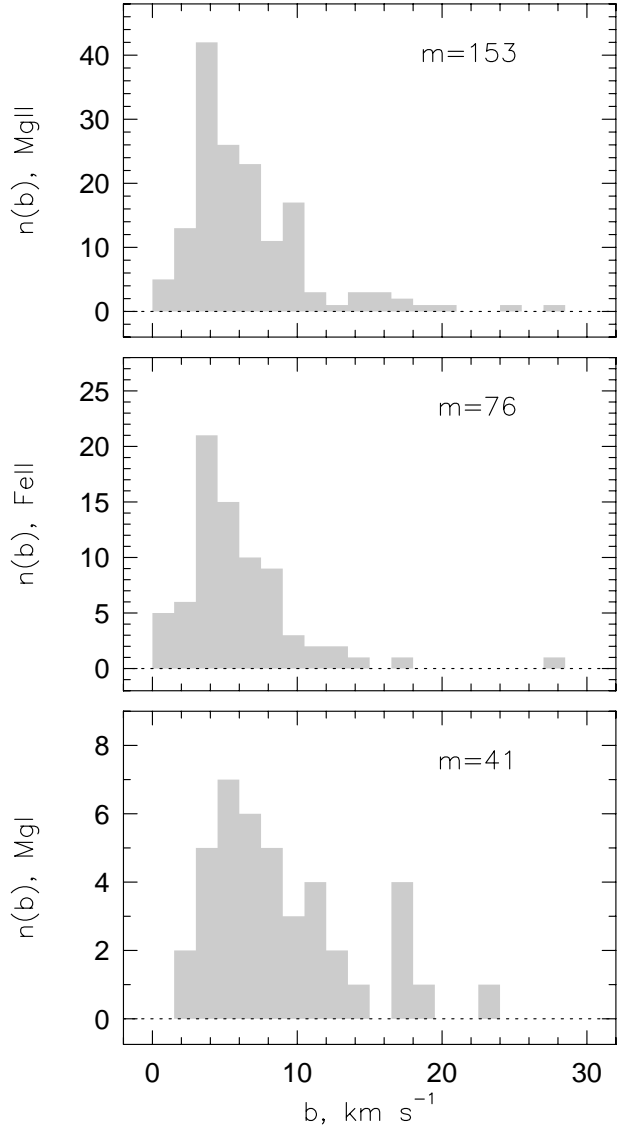}
\figurenum{7}
\caption
{The binned distributions of the VP Doppler $b$ parameters for {\MgII}
(top), {\FeII} (center), and {\MgI} (bottom).  The median Doppler
parameters and the standard deviations of the distributions are 
$\left< b \right> = 5.4\pm4.3$, $5.1\pm4.1$, and $7.7\pm5.1$, for 
{\MgII}, {\FeII}, and {\MgI}, respectively. The number of clouds,
$m$, in each sample is written in each panel.
\label{fig:bdist}}
\end{figure*}

The binned distribution of {\MgII}, {\FeII}, and {\MgI} Doppler $b$
parameters is presented in Figure~\ref{fig:bdist}.  The median Doppler
parameters and the standard deviations of the distributions are
$\left< b \right> = 5.4\pm4.3$, $5.1\pm4.1$, and $7.7\pm5.1$, for
{\MgII}, {\FeII}, and {\MgI}, respectively.  We note that,
statistically, the relative distributions of {\MgII} and {\MgI} are
unphysical; it is difficult to understand how {\MgI} could have
broader lines than {\MgII} on average.

Based upon simulations, the observed distribution peak for {\MgII}
could be shifted to a larger value by $\sim 1$--$2$~{\kms} compared to
the ``true'' distribution.  The true {\MgII} distribution is likely
peaked at $b = 3$--$4$~{\kms}.  Simulations further indicate that the
true distribution may have zero to few clouds with $b \leq 1$~{\kms}.
A lower cut--off $b$ parameter of $1.5$~{\kms} is required to not
overproduce the lowest bin.  However, we note that this could be a
resolution effect, given that lines are unresolved for $b \leq
2.46$~{\kms}.

In $30$~{\kms} resolution data, \citet{pb90} measured a {\MgII} $b$
distribution peaking at $10 \leq b \leq 15$~km~s$^{-1}$, with a
decaying tail at $b \geq 20$~{\kms}.  They suggest that non--thermal
or turbulent motions within the individual clouds are implied because
such large $b$ values correspond to temperatures in excess of
$10^{5}$~K, a temperature too high for {\MgII} to survive if
photoionized.  Our results are suggestive of clouds that are less
turbulent and/or cooler, with temperatures between $30$--$40,000$~K.
In our data, all VP components with $b>10$~{\kms} occur in broad,
partially or fully saturated profiles.

Simulations revealed that the majority of the large $b$ parameters
forming the extended tail to the distribution is an artifact of
component blending.  Approximately $70$\% of the input VP
components were recovered; that is, three of ten components were
typically lost because of blending\footnote{This result is sensitive
to the signal--to--noise ratio of the spectra.  The quoted results are
for signal--to--noise ratios commensurate with the data.}.
As such, some components may have large $b$ parameters in cases where
two components may have been responsible for the absorption
profile, but could not be formally constrained by the fitting
algorithm.   

\subsection{Thermal and Turbulence Broadening}

The  $b$  parameter  of a  given  ion  can  have  both a  thermal  and
non--thermal  component.    This  latter  contribution   to  the  line
broadening could  be due to  internal turbulent motion  or systematic,
bulk motion.  Assuming the non--thermal component can be parameterized
as a Gaussian distribution  preserves the Voigt profile formalism.  We
note  that a  Gaussian non--thermal  motion  in an  isolated cloud  is
difficult   to   understand,   since   it   implies   a   mixing,   or
quasi--convective process  that would be  expected to decay away  in a
cloud dynamical time.

The total $b$ parameter of ion $i$ is written,
\begin{equation}
b^{2}_{i} = b^{2}_{i,therm} + b^{2}_{turb} ,
\end{equation}
where $b_{i,therm}= \sqrt{2kT/m_{i}}$  is the thermal component, which
scales  with the  ion mass  $m$,  and $b_{turb}$  is the  non--thermal
component, which is equal for  all ions.  The thermal and non--thermal
$b$  parameters for  two  ions  of different  mass  are determined  by
solving two equation with two unknowns, which gives
\begin{equation}
b_{turb} = 
   \left[ \frac{b^{2}_{j} - (m_{i}/m_{j}) 
   b^{2}_{i}}{1-(m_{i}/m_{j})} \right] ^{1/2} ,
\label{eq:bturb}
\end{equation}
and 
\begin{equation}
b_{i,therm} = 
   \left[ \frac{b^{2}_{i} - b^{2}_{j}}{1-(m_{i}/m_{j})} \right] ^{1/2} ,
\label{eq:btherm}
\end{equation}  
where  $b_{i}$ and  $b_{j}$ are  the measured  $b$ parameters  for the
lighter and heavier ions,  respectively.  This technique for exploring
the  line  broadening mechanism  has  been  applied  using {\CIV}  and
{\SiIV} at $z\sim2.5$ by \citet{rauch-civ}.

\begin{figure*}[p]
\plotone{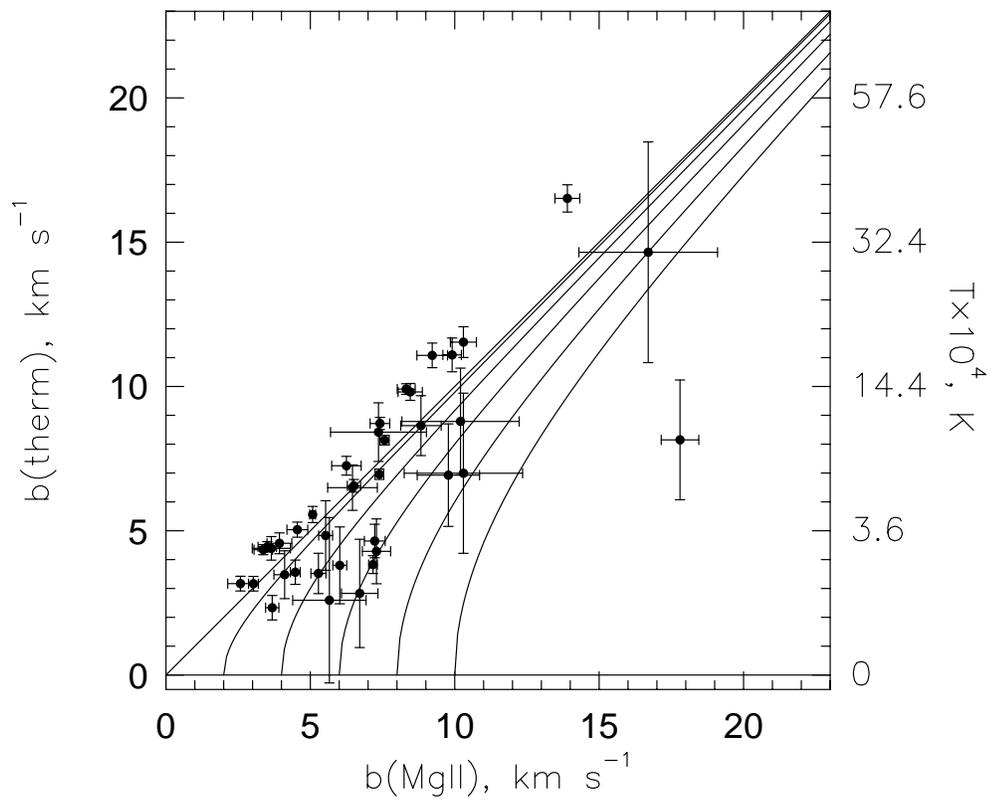}
\figurenum{8}
\caption
{The thermal Doppler $b$ component versus the totel Doppler $b$ for
{\MgII}.  The deconvolution is based upon the {\FeII} total Doppler $b$.
The corresponding temperature (in Kelvin) is provided along the right
hand axis.
\label{fig:turb}}
\end{figure*}

In Figure~\ref{fig:turb}, we present the results of deconvolving the
thermal and non--thermal components to the line broadening of {\MgII}
and {\FeII} components.  The selected data have $b$ parameters with
fractional errors less than 50\% for both {\MgII} and {\FeII}.  The
thermal component is plotted along the vertical and the total $b$
parameter is plotted along the horizontal axis.  The solid curves are
lines of constant non--thermal motion.  The temperature scale for the
thermal component is given on the right hand vertical axis.

Clouds that lie above the $b_{turb}=0$~{\kms} curve have unphysical
{\MgII} to {\FeII} $b$ parameter ratios (recall that the $b$
parameters were fit assuming no constraining relationship between
different ions).  The mean turbulent component of the subsample is
$0.7\pm0.3$~{\kms}.  VP analysis of simulated spectra in which the
clouds are assumed to be thermally broadened, i.e.\ have
$b_{turb}=0$~{\kms}, yielded a mean $b_{turb}$ of $\sim 1$~{\kms}.
This value is consistent with the value obtained for the data and is 
suggestive of thermally broadened clouds.

Furthermore, the results from the simulated spectra showed the same
level of scatter as the observed data plotted on
Figure~\ref{fig:turb}.  As such, it is difficult to claim a
non--thermal component in individual systems on a case--by--case
basis.  Based upon the simulations, it is found that the scatter
decreases with increasing signal--to--noise ratio.  We also note that
even though the multiple {\FeII} transitions offer greater constraints
on the {\FeII} $b$ parameter, the weaker strength {\FeII} data often
have signal--to--noise ratios a factor of $2$--$3$ lower than the
{\MgII} data.
 
The simulation results and the spread of the data about the
$b_{turb}=0$~{\kms} curve provide no compelling statistical evidence
for a non--thermal component to the line broadening of {\MgII}
absorbing gas.  From thermal components, the cloud temperatures are
inferred to be in the range $30,000$--$150,000$~K.  Based upon the
photoionization modeling, it is difficult to understand temperatures
in excess of $60,000$~{K}.  Again, we note that the clouds with the
largest $b$ parameters are due to the blending of components.

\subsection{Cloud--Cloud Velocity Clustering}
\label{sec:velocity}

We computed the velocity two--point clustering function (TPCF) for the
VP components obtained with our HIRES/Keck spectra (these data are
listed in Table~7 of Paper~I).  As computed here, the TPCF, $P(\Delta
v)$, is the probability that any randomly selected pair of VP
components in a system will have a velocity splitting $\Delta v$.

Using  low   resolution  spectra,  \citet{ssb88} used  the TPCF
to show that {\MgII} systems clustered on velocity scales of $\Delta v
\leq  200$~{\kms}.  This was  taken as  evidence that  {\MgII} systems
cluster like field galaxies.  Using moderate resolution spectra ($\sim
30$~{\kms}),  \citet{pb90} found  that a
two    component    Gaussian    model   with    $\sigma_{1}=80$    and
$\sigma_{2}=390$~{\kms} provided a good fit to the {\MgII} TPCF.  They
interpreted  the smaller  width to  reflect the  kinematics  of clouds
bound  within galactic  halos and  the  broader width  to reflect  the
kinematics of galaxy pairs in the field.

\begin{figure*}[p]
\plotone{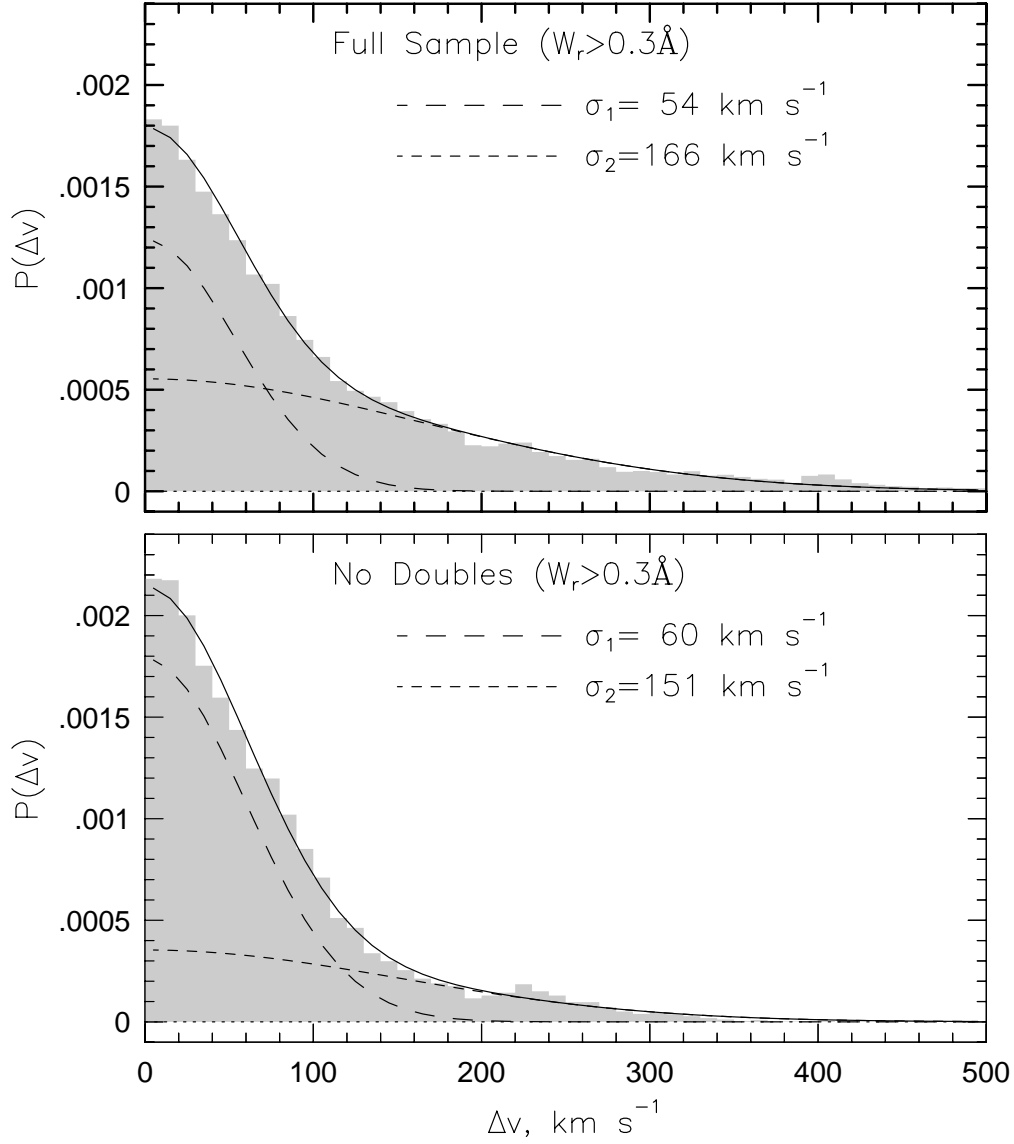}
\figurenum{9}
\caption
{(top) The VP component velocity two--point correlation function
(TPCF) for the full sample.  The distribution is well described by a
double Gaussian function with a narrow dispersion of $\sigma _{1} =
54$~{\kms} and a broader dispersion of $\sigma _{2} = 166$~{\kms}.
--- (bottom) The TPCF for the with double systems removed.  The
narrow dispersion is $\sigma _{1} = 60$~{\kms} and the broader
dispersion is $\sigma _{2} = 151$~{\kms}.
\label{fig:tpcfA}}
\end{figure*}

The  TPCF for  the full  sample  is presented  in the  upper panel  of
Figure~\ref{fig:tpcfA}.  The TPCF  increases sharply
for $\Delta v \leq 100$~{\kms} and has an extended tail out to $\simeq
400$~{\kms}.  Following \citet{pb90}, we
parameterized the TPCF as a two--component Gaussian,
\begin{equation}
P(\Delta v) = 
A_{1}\exp \left( \frac{\Delta v^{2}}{2\sigma ^{2}_{1}} \right) +
A_{2}\exp \left( \frac{\Delta v^{2}}{2\sigma ^{2}_{2}} \right) ,
\end{equation}
where the $A$ and $\sigma$ are the amplitudes and Gaussian widths,
respectively.  We obtained a narrow width of $\sigma _{1} =
56$~{\kms} and a broader width of $\sigma _{2} = 166$~{\kms} for
the full sample.  The relative amplitudes differ by a factor of two,
with the narrow component dominating.

We examined whether the extended tail of the distribution is dominated
by the three kinematically complex systems, S1, S9, and S12 (see
Figure~\ref{fig:data}).  Not only do these systems have a large
kinematic spread, they have roughly twice the {\Lya} absorption
strengths and very strong {\CIV} absorption compared to ``classic''
{\MgII} systems; they are classified as ``double systems''
\citep{archiveII}.

If  the  double systems  are  removed from  the  sample  and the  TPCF
recomputed,  we obtain  the results  presented in  the lower  panel of
Figure~\ref{fig:tpcfA}.  For this ``No Doubles'' sample, we obtained a
narrow  width of  $\sigma _{1}  = 60$~{\kms}  and a  broader  width of
$\sigma   _{2}  =   151$~{\kms}.    The  component   widths  are   not
substantially  different  from the  full  sample  TPCF.  However,  the
amplitude of the broad component is a factor of five smaller than that
of  the narrow  component.   This indicates  that  the double  systems
strongly  govern   the  amplitude  of  the  broad   component  to  the
full--sample  TPCF, but that  the component  still results  because of
``moderate--''       and       ``high       velocity''       kinematic
subsystems\footnote{Kinematic subsystems are  defined in Paper~I.}  in
the ``classic'' {\MgII} absorbers \citep{archiveII}.

\subsection{Evolutionary Clues from Kinematics}

It is difficult to  apply a straight--forward, physical interpretation
to  each of  the  two components  of  the TPCF.   However, based  upon
absorber  models \citep{thesis,kinematicpaper},  the narrow
component  width  may be  proportional  to  the vertical  velocity
dispersion of gas in  galaxies disks (averaged over random inclination
angles).  This  is also consistent  with the notion that  the dominant
kinematic  subsystems  (see Paper~I),  could  arise in  systematically
rotating  disks \citep[e.g.,][]{lb92,kinematicpaper,cv01}.

The data directly constraining this scenario are few in number; in
four out of five edge--on spiral galaxies hosting $W_{r}(2796) \geq
0.3$~{\AA} absorption, the gas kinematics directly trace the outer
extensions of the stellar disk rotation curves \citep{steidelpaper}.
However, the data are difficult to understand in terms of a simple
disk model.  Counter examples are also few; \citet{lanzetta97} and
\citet{ellison02} have each reported an example inconsistent with a
rotating disk scenario.

Clues to the nature of the broader component to the TPCF might be
connected to the redshift evolution of $W_{r}(2796)$.  \citet{ss92}
showed that as the minimum equivalent width of a {\MgII}--absorber
sample is increased, the redshift number density evolution becomes
stronger.  That is, the strongest systems evolve away most rapidly
with decreasing redshift (increasing cosmic time) from $z\simeq 2$ to
$z\simeq 0.3$.  The underlying physical process governing this
evolution remains a mystery, though it has been attributed to either a
reduction in the mean column density of the clouds with time
(chemical, ionization, and/or gas mass evolution) or a reduction in
the velocity spread of the clouds with time (kinematic evolution), or
both.

To explore trends in the kinematics with equivalent width, we computed
a separate  TPCF for samples B, C,  D, and E, with  and without double
systems.   The   results  are  listed   in  Table~\ref{tab:tpcf}.  From
these  distributions,  we see  that  the  nature  of the  cloud--cloud
clustering is strongly connected to the equivalent width.

Including double systems, the velocity dispersion of the narrow
component, $\sigma _{1}$, does not change significantly with
increasing $W_{r}(2796)$, whereas the broad component velocity
dispersion, $\sigma_{2}$, decreases and its relative amplitude,
$A_2/A_1$, increases.  If our sample is taken to be representative of
the population of {\MgII} absorbers, this trend in the TPCF would
suggest that the frequency of highest velocity clouds decreases as
equivalent width is increased (note that the exclusion of the double
systems makes no change to the results for sample B.)  In other words,
in the largest equivalent width systems, the cloud velocities are
distributed more uniformly throughout the full velocity interval of
the profile and clouds at extreme velocities are less frequent.

Excluding double systems, the extended tail in the TPCF vanishes for
samples D and E.  Apparently there are two types of kinematics
characterizing the very strongest equivalent width systems, and they
are extreme to one another.  The first type is exemplified by systems
S1 and S9 (double systems), where the profile morphology is
characterized by a large velocity spread of many clouds representing a
wide range of optical depths.  Systems S5 and S7 (which are DLAs or
near--DLAs) are examples of the second type.  The {\MgII} profile
morphology of DLAs and near--DLAs, in the studied redshift regime, are
characterized by fully saturated, black--bottom absorption with a
moderate kinematic spread of $\sim 60$~{\kms} \citep{archiveII}.

These findings may provide insight into the nature of the differential
equivalent width evolution quantified by \citet{ss92}.  Since the
redshift number density of DLAs apparently does not evolve over the
redshift range 0.5 to 2.0 \citep{rao2000}, those large equivalent
width {\MgII} systems arising in DLAs would also not evolve over this
redshift interval.  Therefore, if the kinematics of the largest
equivalent width systems do fall into the above two categories, and
one type arises primarily in DLAs, then it must be the double systems
undergoing the most rapid evolution.

This would suggest that {\MgII} absorption cross section of the
kinematically higher velocity material is decreasing with time.
However, moderate-- and high--velocity kinematic subsystems with weak
absorption do not contribute significantly to the total equivalent
width in a system.  Thus, the evolution cannot be dominated by a
decrease in the frequency of weak subsystems in double systems like S1
and S12.  {\it It must be an evolution in systems more like S9, where
the optical depth is highly variable and relatively large across the
entire kinematic range of absorption}.  At even larger equivalent
widths, $\simeq 2$~{\AA} and greater, the evolution is likely dominated
by systems with kinematics similar to those studied by \citet{bond}.
A sample of high resolution {\MgII} systems at $z>1$ will be required
to ultimately address these issues.

\subsection{Column Densities and Kinematics}

\begin{figure*}[p]
\plotone{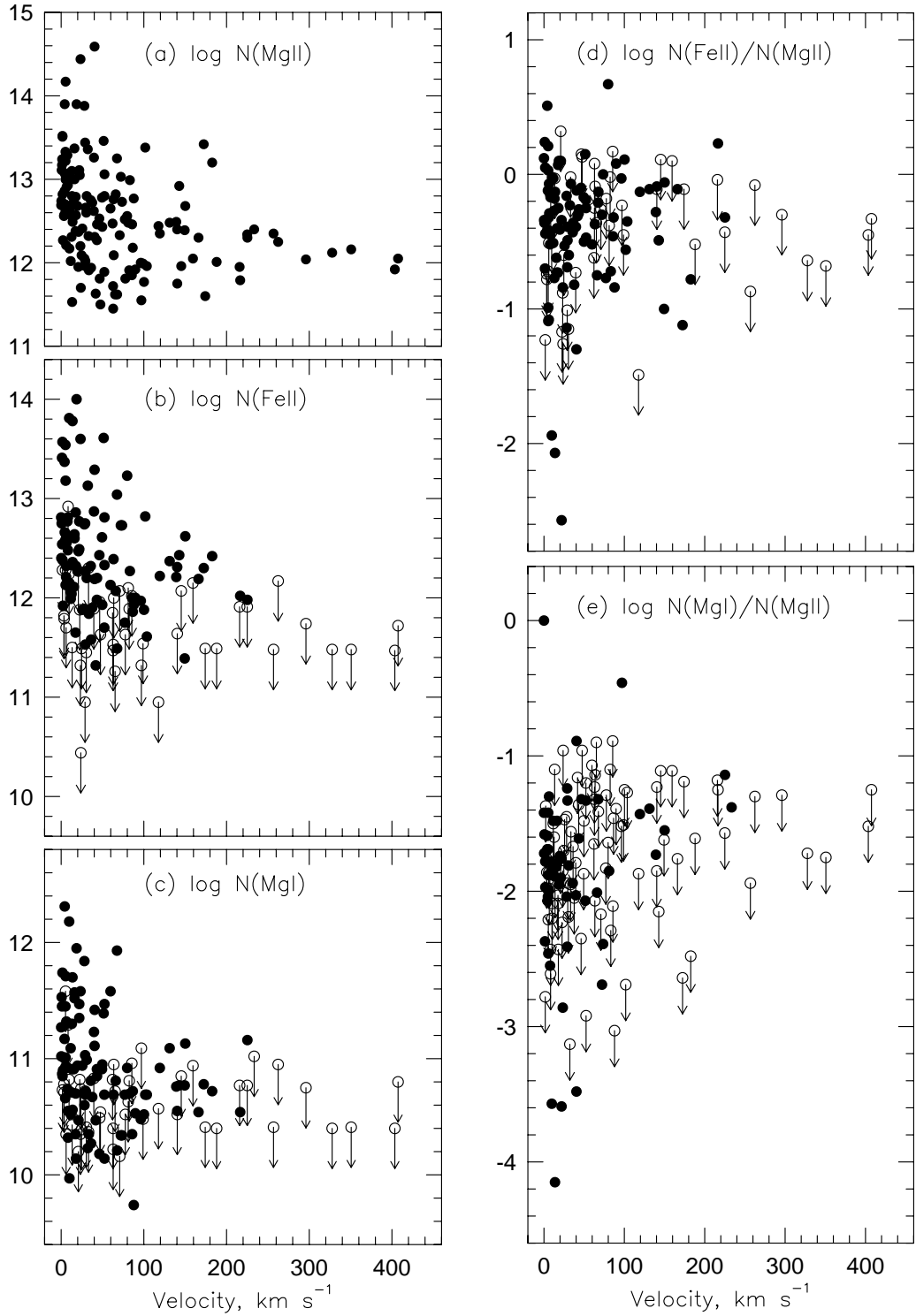}
\figurenum{10}
\caption
{(a,b,c) The logarithmic VP column densities of {\MgII}, {\FeII}, and
{\MgI} as a function of cloud velocity.  --- (d,e) The logarithmic VP
column density ratios of {\FeII} and {\MgI} to {\MgII} as a function
of cloud velocity.  For all panels, the vertical scales are adjusted
for ease of comparison. The error bars are not shown for presentation
purposes; however, the errors and data point types can be viewed in
Figure~\ref{fig:mg2+fe2+mg1}.
\label{fig:Rvsvel}}
\end{figure*}

In Figures~\ref{fig:Rvsvel}$a$--\ref{fig:Rvsvel}$c$, we present the
logarithmic column densities for {\MgII}, {\FeII}, and {\MgI}.  Solid
points are measurements and open points are upper limits.  The
anti-correlations with velocity are induced by the definitions of the
velocity zero point.  The zero points are defined as the optical depth
median of the absorption profiles and virtually all of them are
aligned with the centroid of a dominant kinematic subsystem (see
Paper~I).  The dispersion of the column densities decreases
with velocity, though it is not clear whether the very small
dispersion for $v > 200$~{\kms} is due to small number statistics.

For $\log N({\MgII}) \leq 12.5$~{\cmsq}, the distribution of {\MgII}
column densities is quite flat with velocity, which indicates that low
column density clouds are present at all velocities.  For these
clouds, {\FeII} and {\MgI} are often below the detection threshold of
the data.  This explains the presence of upper limits for {\FeII} and
{\MgI} at all velocities, and most obviously at high velocities.

For our sample, we can at least say that it is rare to find large
column density clouds at large velocities from the dominant kinematic
subsystem (also see Paper~I).  As trivial as this
observation may seem, it has implications for the systematics in the
absorption kinematics and therefore possibly the absorber geometry.
Apparently, finding two large kinematic subsystems in a single
absorber is rare.  These subsystems are comprised of several blended
components, which is suggestive of gas that is part of a coherent
structure with low velocity dispersion, i.e.\ systematic kinematics.
Since there is only a single dominant kinematic subsystem per absorber
and if its kinematics are systematic, the geometry is likely to be
quasi--planar with velocities primarily parallel to the plane.

In Figures~\ref{fig:Rvsvel}$d$ and \ref{fig:Rvsvel}$e$, we present
$\log N({\FeII})/N({\MgII})$ and $\log N({\MgI})/N({\MgII})$ vs.\
cloud velocity.  Interestingly, the dispersion of $\log
N({\FeII})/N({\MgII})$ with velocity is fairly constant with a range
from $-1$ to $+0.2$ out to $v \simeq 200$~{\kms}.  This would suggest
that the range of physical conditions (abundance pattern, dust
depletion, ionization conditions) governing this ratio does not change
dramatically with kinematics.  If, as suggested above, the main
kinematic subsystems are single coherent structures, then it may be
that the higher velocity clouds are ``cloudlettes'' surrounding this
structure.  If so, then these ``cloudlettes'' apparently do not differ
greatly in their {\FeII} to {\MgII} conditions.  Beyond $v\simeq
200$~{\kms} we can only comment that there are no clouds in our sample
with elevated $N({\FeII})/N({\MgII})$.

Since {\MgI} absorption is present primarily at velocities less than
$\simeq 100$~{\kms}, it is less clear if the dispersion of $\log
N({\MgI})/N({\MgII})$ with velocity is decreasing at larger
velocities.  In fact, 65\% of the detected {\MgI} arises within $v <
30$~{\kms} and 80\% arises with $v < 60$~{\kms}.  The range of
$N({\MgI})/N({\MgII})$ for $v < 60$~{\kms}, however, is four orders
of magnitude (excluding the point at $\log N({\MgI})/N({\MgII}) =
0$).  This suggests that the conditions where {\MgI} arises are
probably quite varied.  This is likely an ionization condition
effect.  It could be due to differing cloud densities, ionizing flux, or
to multiphase ionization conditions (see next section).

\section{Photoionization Models}
\label{sec:models}

A grid of Cloudy 90 \citep{ferland} photoionization models were
constructed in order to place constraints on the chemical and
ionization conditions of the VP components.  Several researchers have
applied photoionization models to {\MgII}--selected absorption systems
\citep[e.g.,][]{bs86,bgy87,bergeron94,q1206}.  A useful exposition on
the methods used here has been written by \citet{steidel90}.

In brief, the clouds are assumed to be plane--parallel, constant
density ``slabs'' with an external ionizing flux incident upon one
face of the cloud.  For this configuration, the ionization condition
is quantified using the ionization parameter, $U = n_{\gamma}/n_{\rm
H}$, where $n_{\gamma}$ is the number density of hydrogen ionizing
photons and $n_{\rm H}$ is the number density of hydrogen atoms.  The
quantity $n_{\gamma}$ is dependent upon both the intensity (amplitude
at 1~Ryd) and shape of the ionizing spectrum.  The clouds are given a
pre--defined neutral hydrogen column density, $N({\HI})$.  This
results in well--defined clouds that are indexed by their $U$ and
$N({\HI})$ values.  We produced a grid of model clouds over the range
$-5 \leq \log U \leq -1$ and $15 \leq \log N({\HI}) \leq 20$~{\cc} at
half dex intervals in both quantities.

In these photoionization models,  the column density ratios of various
ions   are  effectively   independent  of   metallicity.    Only  when
metallicity approaches the solar value, resulting in increased cooling
rates, are the ratios modified from those in lower metallicity clouds.
We use 1/10th solar metallicity ($Z  = -1$) for our models.  We also use
a  solar abundance  pattern.   

\subsection{Ionizing Spectrum}

The spectral energy distribution  (SED) and intensity of the ionizing
flux is most uncertain.  We explored both an extragalactic ultraviolet
background (UVB)  source and ``local'' stellar sources.   For the UVB,
we  used the spectra  of \citet{hm96}.  These
spectra have only  slightly different SEDs for $z=1$  and $z=0.5$ with
normalizations  at 1~Ryd  of  $\log  \nu f_{\nu}  =  -5.2$ and  $-5.6$
erg~s$^{-1}$~cm$^{-1}$,   respectively.   We   find   that  there   is
negligible  difference in  the column  density ratios  of  {\FeII} and
{\MgI} to {\MgII} for cloud models ionized by these two UVB spectra.
For the following discussions, we use the $z=1$ SED.

Assuming no shielding of the UVB, stellar radiation must dominate over
the UVB before the stellar SEDs begin to alter column density ratios
in cloud models.  The ionization potentials to destroy {\MgII} and
{\FeII} are slightly greater than 1~Ryd, whereas the ionization
potential to destroy {\MgI} is slightly below.  As shown in Appendix~B
of \citet{cl98}, only O and B stars are capable of modifying the UVB
in this energy range for astrophysically reasonable stellar number
densities.  They also reported that their results were insensitive to
stellar surface gravity.  Thus, we limited our exploration to stars
with $T=30,000$~K, to represent B stars, and with $T=50,000$~K, to
represent O stars.  Following \citet{cl98}, we used the solar
metallicity stellar models produced by \citet{kuruzc91}.

As with the UVB grids, we produced a grid of model clouds over the
range $-5 \leq \log U \leq -1$ and $15 \leq \log N({\HI}) \leq
20$~{\cc} at half dex intervals in both quantities.  We also varied
the $\log \nu f_{\nu}$ of stars at 1~Ryd over the range $-7$ to $-3$
erg~s$^{-1}$~{\cmsq} at 1~dex intervals. We produced a separate grid
for B and O stars, respectively.  For the stellar SEDs, there is a
sharp edge at 1~Ryd with a relatively flat energy distribution at
energies just greater than 1~Ryd.  Thus, for subsequent discussion,
we focus on the normalization of the stellar grids at 1.2~Ryd, which
is very close to the ionization potentials to destroy {\MgII} and
{\FeII}.  For reference, the $z=1$ UVB at 1.2 Ryd has $\log \nu
f_{\nu} = -5.3$ erg~s$^{-1}$~{\cmsq}.

\subsection{Limits on Stars and Their Distances}

Following the formalism of \citet{cl98},  we derive the minimum
number of O, B~I, and B~V stars in a confined region at a distance of
$R$ kiloparsecs, that are required to influence the ionization
conditions to be
\begin{equation}
\log N_{\ast}({\rm O}) = 0.1 + 2\log R ,
\label{eq:ostars}
\end{equation}
\begin{equation}
\log N_{\ast}({\rm B~I}) = 0.0 + 2\log R , 
\label{eq:bistars}
\end{equation}
\begin{equation}
\log N_{\ast}({\rm B~V}) = 0.8 + 2\log R ,
\label{eq:bvstar}
\end{equation}
at $z=1$.  The constants are reduced by 0.4 for $z=0.5$.  These
constants were computed using the flux level of the UVB SED at
1.2~Ryd, effectively at the ionization potentials of {\MgII} and
{\FeII}.  To recompute $N_{\ast}$ for a known stellar flux, $\nu f_{\nu}$ at
1.2~Ryd, one simply adds $5.3 + \log \nu f_{\nu}$.  

\begin{figure*}[p]
\plotone{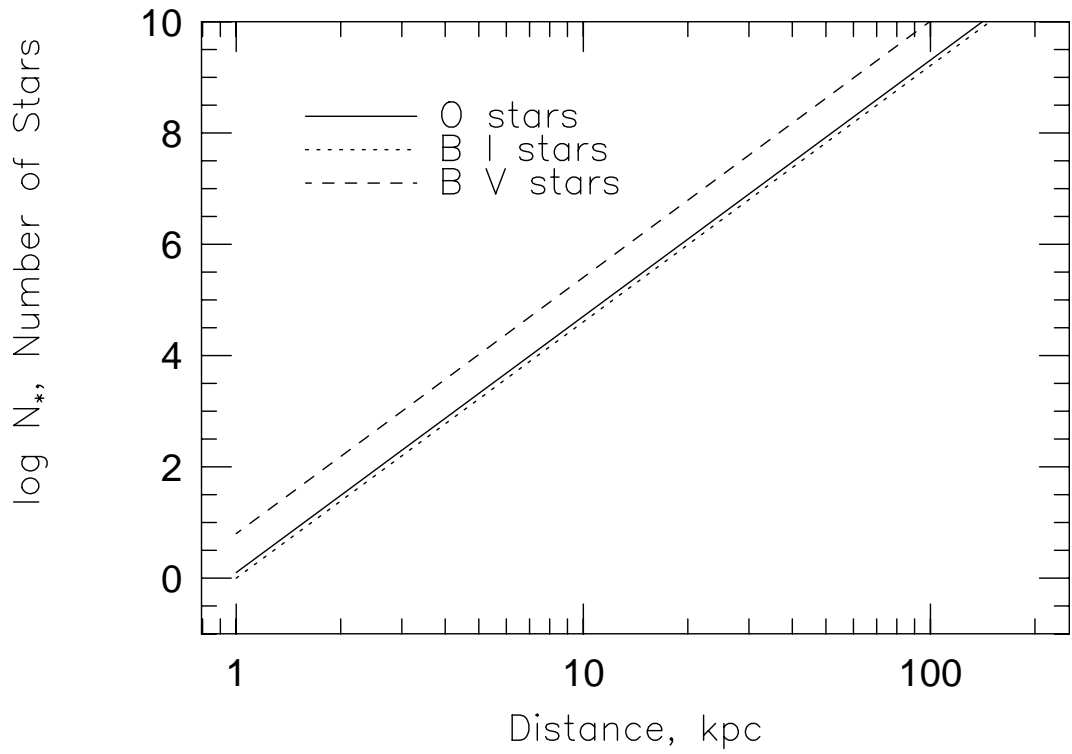}
\figurenum{11}
\caption
{The minimum number of stars required at $z=1$ to dominate the
extragalactic ultraviolet background radiation as a source of
photoionization.  The curves are 0.4 dex lower at $z=0.5$.  Three
types of stars are shown: O stars (solid), B I stars (short dash), and
B V stars (long dash).
\label{fig:nstars}}
\end{figure*}

We have assumed that the photon escape fraction from the stellar
environment is 100\%.  Under the assumption of a grey
opacity, $\log N_{\ast}$ would be increased by $|\log f_{esc}|$, where
$0 < f_{esc} \leq 1$.  In Figure~\ref{fig:nstars}, we plot $\log
N_{\ast}$ vs.\ $\log R$ for the three stellar types at $z=1$.  We
apply these results below.

\subsection{{\FeII} to {\MgII} Ratio}

\begin{figure*}[p]
\plotone{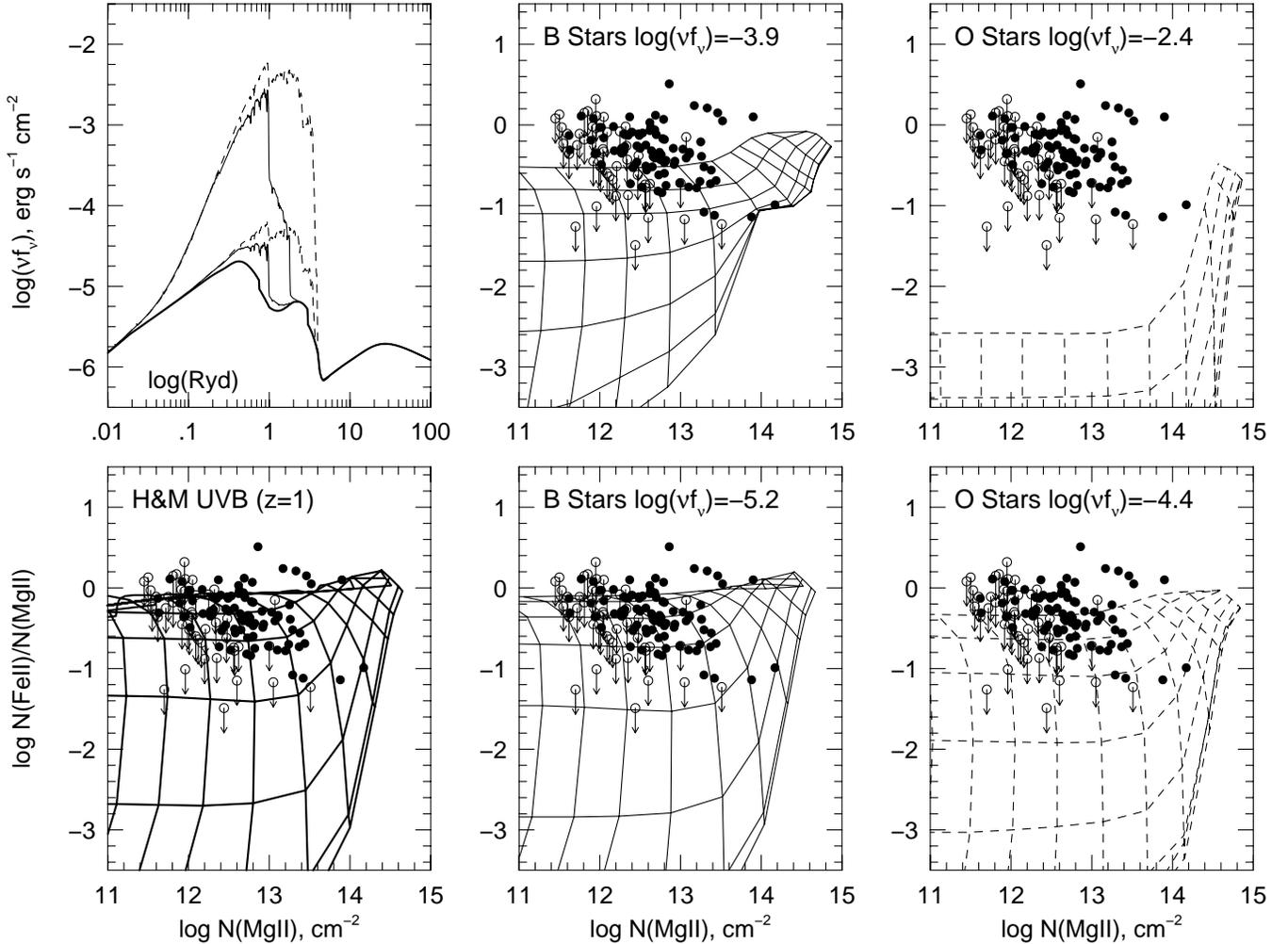} 
\figurenum{12}
\caption
{ (upper left) Spectral energy distributions (SEDs) of the ionizing
flux for the photoionization models.  The thick solid curve is the
Haard \& Madau UV background for $z=1$.  The thin solid curves are
$T=30,000$~K Kuruzc stellar models (B stars) with normalization $\log
(\nu f_{\nu}) = -3.9$ and $-5.2$ erg~s$^{-1}$~{\cmsq}.  The thick
dashed curves are $T=50,000$~K Kuruzc stellar models (O stars) with
normalization $\log (\nu f_{\nu}) = -2.4$ and $-4.4$
erg~s$^{-1}$~{\cmsq}.  The photoionization model grids are shown in
the remaining panels for $\log N({\FeII})/N({\MgII})$ vs.  $\log
N({\MgII})$ as labeled with the data points overplotted.  The error
bars are not shown for presentation purposes; however, the errors and
data point types can be viewed in Figure~\ref{fig:mg2+fe2+mg1}.  The
vertical curves are constant neutral hydrogen column density starting
with $\log N({\HI}) = 20$~{\cmsq} on the right and decreasing by 0.5
dex moving leftward on the diagrams.  The horizontal curves are
constant ionization parameter, starting with $\log U = -5$ at the top
and decreasing by 0.5 dex moving downward on the diagrams. 
(See text for details.)
\label{fig:cloudy-mg2-fe2}}
\end{figure*}

In Figure~\ref{fig:cloudy-mg2-fe2}, we show results of the
photoionization models for the {\FeII} to {\MgII} column density
ratios overplotted on the data.  Six panels are shown.  The upper left
panel shows the SEDs for the assumed ionizing continuum.  The thick
solid curve is the Haardt \& Madau UVB at $z=1$.  The thinner solid
curves are the SEDs of $T=30,000$ (B--type) stars for flux levels
$\log (\nu f_{\nu}) = -3.9$ and $-5.2$ erg~s$^{-1}$~cm~$^{-2}$ at 1.2
Ryd.  The dashed curves are the SEDs of $T=50,000$ (O--type) stars for
flux levels $\log (\nu f_{\nu}) = -2.4$ and $-4.4$
erg~s$^{-1}$~cm~$^{-2}$ at 1.2 Ryd.  These normalizations were chosen
because the lower flux grids illustrate minor modifications and the
higher flux grids represent significant modifications to the UVB
models.  The lower left panel shows the UVB photoionization grid
superimposed upon the data.  The mostly vertical curves are constant
logarithmic neutral hydrogen column density, $\log N({\HI})$, and the
mostly horizontal curves are constant logarithmic ionization
parameter, $\log U$.  The column density contours start at $\log
N({\HI}) = 20$~{\cmsq} and decrease by 0.5 dex intervals toward the
left.  The ionization parameter contours start at $\log U = -5$ and
increase by 0.5 dex intervals downward.  The four remaining panels
show the grids for the stellar ionizing continua.  The grid contours
follow the same pattern.

The UVB model is consistent with a larger majority of
the measured $N({\FeII})/N({\MgII})$ ratios.  There are, however,
several clouds that have measured ratios of $\log
N({\FeII})/N({\MgII}) > 0$ and lie above the model grid.  Adjusting
the $\alpha$--group to iron--group abundance pattern, [$\alpha$/Fe],
in the models is equivalent to ``sliding'' the model grid parallel to
the $N({\FeII})/N({\MgII})$ axis.  However, astronomical ratios range
from $[\alpha/{\rm Fe}] = 0$ (solar) for $-1 \leq Z \leq 0$ to
$[\alpha/{\rm Fe}] = +0.5$ ($\alpha$ enhancement) for $Z < -1$
\citep[e.g.,][]{jtlpasp}.  The model grid would slide downward 0.5 dex
for the $\alpha$ enhancement pattern.  Since we assumed a solar
pattern for the models the grid cannot be elevated further.  Thus, it
is not possible to model these these clouds by assuming a standard
non--solar [$\alpha$/Fe].  It is also not possible to simply assume a
standard dust depletion pattern.  In warm interstellar medium clouds,
iron depletes more readily than does magnesium; iron and magnesium
have depletion factors of $\delta = -1.4$ and $-0.7$, respectively
\citep{jtlpasp}.  Assuming a standard dust depletion would slide the
model grid downward, aggravating the problem.

Some of the elevated column density ratios could be an artifact of the
VP fitting parameterization.  At Keck/HIRES resolution, an unresolved
{\MgII} line saturates (core intensity of zero) at $\log N \simeq
13$~{\cmsq}. In this regime of $N({\MgII})$, the VP fitting
simulations of blended absorption lines yielded roughly 10\% by number
VP components with similarly elevated $N({\FeII})/N({\MgII})$ ratios.
It is thus plausible that a fair fraction of those clouds with $\log N
\simeq 13$~{\cmsq} have elevated $N({\FeII})/N({\MgII})$ ratios as an
artifact of the VP fitting process.

However, this explanation cannot be invoked for those clouds with
$\log N({\MgII}) < 13$~{\cmsq}.  The VP fitting simulations in this
regime of $N({\MgII})$ do not yield any components with elevated
$N({\FeII})/N({\MgII})$ ratios.  Interestingly, in 	our sample, 15
clouds with $\log N({\MgII}) < 13$~{\cmsq} have $N({\FeII})/N({\MgII})
> 0$.  It is worth noting that the ratios in these clouds hold some
similarity to the ``iron--rich'' weak {\MgII} systems \citep{weakII}.
Though the global environments of the clouds in these strong systems
may be very different than those of the weak systems, the physical
processes that give rise to their abundance ratios \citep[see
discussion in][]{weakII} may be one and the same local to the clouds.

Using {\FeII} and {\MgII} as a diagnostic of the presence of stellar
flux is not very useful because of the large degeneracy between the
column density ratios predicted for the different spectral shapes.
The stellar SEDs tend to yield a maximum {\FeII} to {\MgII} ratio that
decreases as the stellar flux is increased.  This is especially true
for the O stars.  It is not ruled out that B stars can be contributing
to the stellar flux for many of the model clouds illustrated.  For B~I
stars, there would need to be $\sim 40,000~(100,000)$ in number to
make a minimum discernible contribution at a distance of
$R=10~(20)$~kpc from the clouds (see Equation~\ref{eq:bistars} and
Figure~\ref{fig:nstars}).

\subsection{{\MgI} to {\MgII} Ratio}

\begin{figure*}[p]
\plotone{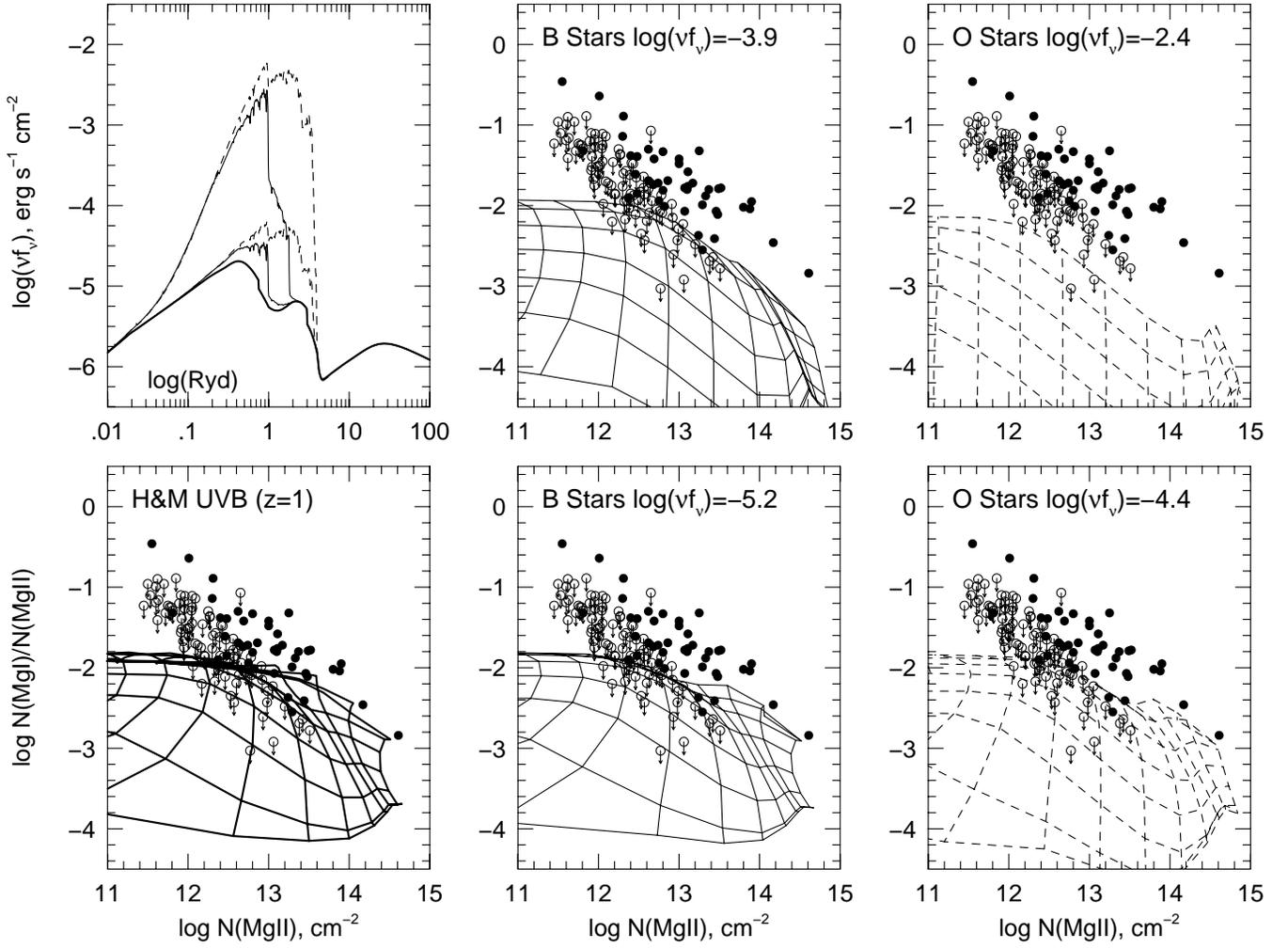} 
\figurenum{13}
\caption
{The same as for Figure~\ref{fig:cloudy-mg2-fe2} but for the
logarithmic VP column density ratio of {\MgI} to {\MgII} as a function
of {\MgII}.  (See text for details.)
\label{fig:cloudy-mg2-mg1}}
\end{figure*}

In   Figure~\ref{fig:cloudy-mg2-mg1},   we   show   results   of   the
photoionization models for the  {\MgI} to {\MgII} column density ratios
overplotted on  the data.  The  photoionization grids are the  same as
described for the  six panels of Figure~\ref{fig:cloudy-mg2-fe2}.  The
{\MgI}  balance is  complicated by  both a  dielectronic recombination
process and a charge exchange reaction with {\HII}.  We compared model
clouds for which these two mechanisms were included to those for which
the mechanisms were ``turned  off''.  In the relatively narrow density
regime and temperature regime of the grids, the inclusion or exclusion
of these processes did not  influence the {\MgI} to {\MgII} ionization
fraction.

Illustrated   in   Figure~\ref{fig:cloudy-mg2-mg1}   are   two   major
discrepancies between  the data  and the models.   

First, there is no single--phase photoionization model presented that
can explain clouds having $\log N({\MgI})/N({\MgII}) > -1.8$.  Though
the results presented here are statistical and involve comparing the
data to a grid of models, we note that even detailed photoionization
modeling involves struggles to reproduce observed {\MgI} to {\MgII} column
density ratios \citep{q1206,ding-q1634,rauch-mgi}.

\citet{ding-q1634} argue that the  majority of the
{\MgI} gas arises  in a lower temperature ($T  \sim 500$~K), virtually
non--turbulent, higher density phase than does the bulk of the {\MgII}
absorption.  That is, the Doppler  $b$ parameters of the {\MgI} clouds
are,  in reality,  significantly smaller  ($b \simeq  1$~{\kms}), than
those of the bulk of the {\MgII} absorption ($b \simeq 5$~{\kms}).  If
Ding  \etal are  correct,  then ``unphysical''  observed large  column
density  ratios are  an artifact  of  VP fitting,  which  does not
enforce  any physical connection  between the  {\MgI} and  {\MgII}
gas.  In  this scenario, much  of the {\MgII}
absorption would  arise in a separate  phase, warmer than  that of the
{\MgI}.    If  so,   then  the   single--phase  models   presented  in
Figure~\ref{fig:cloudy-mg2-mg1}  do  not apply  to  the a  substantial
fraction of the {\MgI} data.

To avoid a high density solution, \citet{rauch-mgi}
suggest a large $N({\MgI})/N({\MgII})$ ratio in the $z=0.5656$ system
toward Q$2237+0305$ may indicate a large $N({\HI})$, on
the order of a DLA.  This cannot be the case for the majority
of the large ratio clouds studied here because (1) it would imply too
large a redshift path density for DLAs in this redshift range
\citep{rao2000}, and (2) only three systems (S2, S5, and S7) are
known to have near--DLA or DLA hydrogen column densities
\citep{archiveI,archiveII}.  

The   second  discrepancy  between   the  data   and  the   models  in
Figure~\ref{fig:cloudy-mg2-mg1}  is that no  model is  consistent with
the observed anti--correlation between $\log N({\MgI})/N({\MgII})$ and
$\log   N({\MgII})$.     This   slope   is    $\Delta   \log
[N({\MgI})/N({\MgII})]/ \Delta \log N({\MgII})  = -0.73$ over the range
$11.5  \leq \log  N({\MgII}) \leq  14.5$~{\cmsq}.   

We searched for other correlations and anti--correlation between the
{\MgI} VP parameters and the {\MgII} VP parameters to see if the
anti--correlation between $\log N({\MgI})/N({\MgII})$ and $\log
N({\MgII})$ could be induced by systematics in the VP fitting.  If
$b({\MgI})/b({\MgII}) > 1$ for the majority of the clouds [an
unphysical condition unless {\MgI} is associated with gas more
turbulent than is {\MgII}], and/or if this ratio were found to
increase as $N({\MgII})$ increases, this could induce the observed
anti--correlation.  However, accounting for uncertainties, all
$b({\MgI})/b({\MgII})$ are consistent with unity and there is no
correlation between $b({\MgI})/b({\MgII})$ and $N({\MgII})$.  There is
also no correlation between $b({\MgI})/b({\MgII})$ and
$N({\MgI})/N({\MgII})$.  As such, there appears to be no obvious VP
fitting correlation that is inducing the observed anti--correlation
between $\log N({\MgI})/N({\MgII})$ and $\log N({\MgII})$.

Interestingly,  the curves  of constant  ionization parameter  for the
highest $\log \nu  f_{\nu}$ stellar models run parallel  to the slope,
$\Delta \log [N({\MgI})/N({\MgII})]/ \Delta \log N({\MgII})$, but they
underpredict the {\MgI}  to {\MgII} column density ratio  by roughly 1
dex.   Also note, however,  that  the stellar  fluxes  for the  illustrated
models require an unreasonable number of ionizing stars; at a distance
of  20~kpc, roughly  $4 \times  10^{5}$ O  stars and  roughly $2\times
10^{6}$ B~I stars are required.

On the other hand,  if the metallicity of the  model clouds were increased
about 1.5 dex, which  would effectively move the photoionization grids
directly      to     the      right     on      the      panels     in
Figure~\ref{fig:cloudy-mg2-mg1},  then  the   model  ratios  could  be
consistent  with the clouds  having $\log  N({\MgII})  \geq 13.5$~{\cmsq}.
However, models of  the lower {\MgII} column density  clouds could not
be  made consistent with  the data.   We also  note that  the required
metallicity  would be  0.5 dex  greater than  solar (this  argument is
independent of abundance pattern).

There appear to be two real discrepancies between the observation and
single--phase ionization models: (1) single--phase models are not
consistent with $\log N({\MgI})/N({\MgII}) > -1.8$; and (2)  the
models cannot explain the
observed anti--correlation between $\log N({\MgI})/N({\MgII})$ and
$\log N({\MgII})$.  As \citet{ding-q1634} point out for two--phase
models, each phase can have $\log N({\MgI})/N({\MgII}) < -1.8$ while
the combined absorption from these two phases can be made consistent
with the total observed absorption strengths (equivalent widths) of
{\MgI} and {\MgII}, respectively.  Further exploration of the
phenomenological and physical conditions governing the {\MgI} to
{\MgII} absorption strengths is beyond the scope of this paper and is
planned for future study.

\section {Conclusions}
\label{sec:conclude}

We have performed an analysis on the Voigt profile fitting results
reported by \citet{cv01} for 23 {\MgII}--selected quasar absorption
line systems.  The quasar spectra were observed with the HIRES instrument
\citep{vogt94} on the Keck~I telescope and have a velocity resolution
of $\sim 6$~{\kms}.  For most of the {\MgII} systems, several {\FeII}
transitions and the {\MgI} $\lambda 2852$ transition were available
for study.  Further details on the data and the Voigt profile fitting
have been presnted by \citet{thesis} and \citet{cv01}.

In general, we have examined both the statistical properties of the
number of components, and the component column densities, Doppler
parameters, and velocities.  Where appropriate, we have compared our
findings with identical analyses on simulated data in order to
ascertain possible statistical bias or systematics.  These simulation
results have provided insights into the degree to which an
observational result is or is not an artifact of the Voigt profile
fitting formalism.  Further details of the simulations can be found in
\citet{thesis}.

The statistical results of the VP analysis are summarized as follows:

(1) The  {\FeII} and {\MgII}  column densities  are correlated  at the
9~$\sigma$  level.    A  least--squares  linear   fit  yielded  $\log
N({\FeII}) =  0.73 \log N({\MgII}) +  3.0$.  A fit to  the {\MgI} column
densities yielded $\log  N({\MgI}) = 0.45 \log N({\MgII})  + 5.2$.  There
is a  5~$\sigma$ anti--correlation of the {\MgI}  column density with
the ratio of  the {\MgI} to {\MgII} column  densities.  

(2) The column density distributions for {\MgII}, {\FeII}, and {\MgI}
were parameterized with a power--law of the form $f(N) \propto
N^{-\delta}$.  The maximum likelihood power--law indices were found to
be $\delta \simeq 1.6, 1.7$, and $2.0$, respectively.  Within
1~$\sigma$ uncertainties, the {\FeII} and {\MgII} column density
indices are consistent with one another.  However, the {\MgI} index is
significantly steeper than those for {\FeII} and {\MgII}.  There is a
5~$\sigma$ anti--correlation between the ratio $N({\MgI})/N({\MgII})$
and $N({\MgII})$ which follows $\log N({\MgI})/N({\MgII}) = -0.73 \log
N({\MgII}) + 7.6$.

(3) The  modes  of  the  Doppler parameter  distributions  were  $\sim
5$~{\kms}  for {\MgII}  and {\FeII}  and $\sim  7$~{\kms}  for {\MgI}.
Based  upon   simulations,  the  true  distribution   modes  could  be
1--2~{\kms} smaller.   The distribution tails  at $b >  10$~{\kms} are
usually due  to blends in  partially or fully saturated  profiles.  We
found that  the clouds are consistent with  being thermally broadened,
with  most common  temperatures  in the  30--40,000~K  range.  

(4) We fitted a two--component Gaussian model to the two--point
correlation function (TPCF) and found velocity dispersions of
$54$~{\kms} and $166$~{\kms}, with the narrow component having twice
the amplitude of the broader component.  An analysis of the
subsamples, demarcated by equivalent width, revealed that the clouds
are distributed more uniformly across the full velocity interval of an
absorption profile as equivalent width increased above $W_{r}(2796) >
0.6$~{\AA}.  In the largest equivalent width systems, extreme velocity
clouds are less frequent than for systems with $ 0.3 < W_{r}(2796)
\leq 0.6$~{\AA}.

(5) A grid of photoionization models, in $\log U$ and $\log N({\HI})$,
were compared to the {\FeII} to {\MgII} and {\MgI} to {\MgII} column
density ratios.  For the most part, the ratios are consistent with
being photoionized by the ultraviolet extragalactic ionizing
background.  Stellar radiation has a difficult time producing clouds
with $\log N({\FeII})/N({\MgII}) \simeq 0$.  As the relative
contribution of the stellar flux increases, the maximum {\FeII} to
{\MgII} column density ratio allowed by the models decreases.  A small
number of clouds are measured to have $\log N({\FeII})/N({\MgII}) >
0$, and there are no models that can produce this ratio.  Similar
conclusions hold for the {\MgI} to {\MgII} column density ratios.  The
majority of the clouds have $\log N({\MgI})/N({\MgII}) > -1.8$, which
cannot be made consistent with single--phase photoionization models.
It is possible that the {\MgI} is arising in a separate, colder phase
with a temperature of several hundred degrees.

Here, we briefly take the opportunity to recap what has been observed
and can be inferred about the absorber kinematics, geometry, and
kinematic evolution.  The {\MgII} absorbers break into kinematic
subsystems, comprised of either a blend of components or a single
component.  In almost all systems there is a single dominant subsystem
with $W_{r}(2796) \simeq 0.2$~{\AA} (on average).  There are no
examples of systems having two dominant subsystems with a substantial
velocity spearation.  The dominant subsystem are comprised of clouds
that are close in velocity and result in blended profiles.
Kinematically, the ``minor'' subsystems either all lie to positive
velocities or all lie toward negative velocities with respect to the
dominant subsystem.  These facts are suggestive of systematic
kinematics and possibly a planar geometry for the dominant subsystem.

For the largest equivalent width systems (sample E), there appear to
be two distinct absorption profile morphologies (i.e.\ optical depth 
as a function of velocity).  These systems have been classified as
either ``double'' systems or DLA/{\HI}--rich systems
\citep{archiveII}.  The double systems are characterized by a large
kinematics spread (greater than 200~{\kms}) and a high variable optical
depth across the majority of the profile (double systems with smaller
equivalent widths are characterized by weak absorption at the larger
velocities).  The DLA/{\HI}--rich systems are characterized by a
moderate velocity spread ($\sim 60$~{\kms}) with and a black--bottomed,
saturated profile.

Now, also consider the following two facts: (1) The population of the
largest systems, with $W_{r}(2796) > 1.0$~{\AA}, evolves away the most
rapidly compared to all {\MgII} systems in the redshift range $z=2$ to
$z=0.3$ \citep{ss92}; and (2) The redshift number density of DLA
systems is apparently {\it not\/} evolving over the studied redshift
range \citep{rao2000}.  Since the DLAs appear to have a distinct
kinematics characterized by black--bottomed saturation and a moderate
velocity spread, we can infer that this particular {\MgII} kinematic
morphology is not evolving.  Thus, it must be the double
systems that are evolving away with redshift.  We further note that it
must be the subset of double systems, like S1 and S9, which are
characterized by large variations in optical depth across the majority
of their velocity spread, that are dominating the evolution.  This is
because the weak higher velocity absorption in double systems, like
S12, do not contribute significantly to the total equivalent width of
the system.   Evolution in these weak subsystems at higher velocities
cannot be contributing to the observed evolution.

Thus, there appears to be an evolving population of very strong
{\MgII} absorbers with complex kinematics that does not include DLAs.
They must be Lyman--limit systems.  It is an interesting question to
ask if these systems are connected with the star formation history of
their host galaxies.  Is there a population of post--bursting galaxies
that fade at $z\sim 2$ but generate kinematically complex, highly
spread out {\MgII} absorbers that passively evolve away to $z \sim 1$?
Or, do these systems trace the ``smooth'' accretion of enriched gas
from the intergalactic medium onto galaxies at higher redshifts that
becomes less dominant at lower redshifts \citep[e.g.,][]{murali02}.

\acknowledgements

Support for this work was provided by NASA (NAG 5--6399), the
California Space Institute (CS--194), Sigma--Xi (Grants in Aid of
Research), and by the NSF (AST 96--17185).  We thank D. Schneider for
helpful discussions about wholesale simulations of spectral analysis.



\begin{deluxetable}{lcccccc}
\tablewidth{0pc}
\tablecolumns{7}
\tablecaption{Samples}
\tablehead
{
\colhead{QSO} &
\colhead{$z_{abs}$} &
\colhead{ID} &
\colhead{Sample B} &
\colhead{Sample C} &
\colhead{Sample D} &
\colhead{Sample E} \\
\colhead{ } &
\colhead{ } &
\colhead{ } &
\colhead{$0.3-0.6$~{\AA}} &
\colhead{$0.6-1.0$~{\AA}} & 
\colhead{$\geq 0.6$~{\AA}} &
\colhead{$\geq 1.0$~{\AA}} 
 }
\startdata
$0002+051$ & $0.85139$ & S1  &   &   & X & X \\
$0117+213$ & $0.57640$ & S2  &   & X & X &   \\
$0117+213$ & $1.04797$ & S3  & X &   &   &   \\
$0420-014$ & $0.63300$ & S4  &   & X & X &   \\
$0454+039$ & $0.85957$ & S5  &   &   & X & X \\
$0454+039$ & $1.15325$ & S6  & X &   &   &   \\
$0454-220$ & $0.47441$ & S7  &   &   & X & X \\
$0454-220$ & $0.48334$ & S8  & X &   &   &   \\
$0823-223$ & $0.91102$ & S9  &   &   & X & X \\
$1101-264$ & $0.35900$ & S10 & X &   &   &   \\
$1148+384$ & $0.55336$ & S11 &   & X & X &   \\
$1206+459$ & $0.92760$ & S12 &   & X & X &   \\
$1222+228$ & $0.66805$ & S13 & X &   &   &   \\
$1241+176$ & $0.55048$ & S14 & X &   &   &   \\
$1248+401$ & $0.77296$ & S15 &   & X & X &   \\
$1254+044$ & $0.51939$ & S16 & X &   &   &   \\
$1254+044$ & $0.93423$ & S17 & X &   &   &   \\
$1317+274$ & $0.66005$ & S18 & X &   &   &   \\
$1421+331$ & $0.90287$ & S19 &   &   & X & X \\
$1421+331$ & $1.17261$ & S20 & X &   &   &   \\
$1634+706$ & $0.99024$ & S21 & X &   &   &   \\
$2128-123$ & $0.42973$ & S22 & X &   &   &   \\
$2145+064$ & $0.79078$ & S23 & X &   &   &   \\
\enddata
\label{tab:samples}
\end{deluxetable}


\begin{deluxetable}{lccccccccc}
\tablewidth{0pc}
\tablecolumns{10}
\tablecaption{TPCF Parameters}
\tablehead
{
\colhead{Sample} &
\colhead{$A_1$} & 
\colhead{$\sigma_1$} & 
\colhead{$A_2/A_1$} &
\colhead{$\sigma_2$} & 
\colhead{ } &
\colhead{$A_1$} & 
\colhead{$\sigma_1$} & 
\colhead{$A_2/A_1$} & 
\colhead{$\sigma_2$} \\
\colhead{ } &
\colhead{$(\times 10^{-3})$ } & 
\colhead{\kms} & 
\colhead{ } &
\colhead{\kms} & 
\colhead{ } &
\colhead{$(\times 10^{-3})$} & 
\colhead{\kms} & 
\colhead{ } & 
\colhead{\kms}
}
\startdata
  & \multicolumn{4}{c}{With Doubles} & & \multicolumn{4}{c}{Without Doubles} \\
\cline{2-5} \cline{7-10}
Full & 1.25 & 54 & 0.45 & 166 & & 1.80 & 60 & 0.20 & 151   \\
B    & 1.51 & 27 & 0.83 &  93 & & 1.51 & 27 & 0.83 &  93   \\
C    & 1.18 & 55 & 0.33 & 273 & & 1.46 & 58 & 0.23 & 195   \\
D    & 0.91 & 66 & 0.55 & 201 & & 1.66 & 81 & \nodata & \nodata \\
E    & 0.78 & 67 & 0.77 & 154 & & 1.77 & 84 & \nodata & \nodata \\
\enddata
\label{tab:tpcf}
\end{deluxetable}


\end{document}